\documentclass[12pt,a4paper]{article}

\usepackage{latexsym}
\usepackage{epsf}
\usepackage{graphicx}
\def\be{\begin{equation}}
\def\ee{\end{equation}}
\def\bea{\begin{eqnarray}}
\def\eea{\end{eqnarray}}
\def\by{\left(\begin{array}}
\def\ey{\end{array}\right)}

\def\slash#1{\setbox0=\hbox{$#1$}#1\hskip-\wd0\dimen0=5pt\advance
       \dimen0 by-\ht0\advance\dimen0 by\dp0\lower0.5\dimen0\hbox
         to\wd0{\hss\sl/\/\hss}}

\newcommand{\ve}[1]{\mbox{\boldmath$#1$}}

\begin{document}

\begin{center}
\Large{Pion Mass Shift and the Kinetic Freeze Out Process} 
\end{center}

\begin{center}

\small {Sven Zschocke$^{1}$, Laszlo P. Csernai$^{1,2}$} 
\end{center}

\footnotesize{
\begin{center}
$^{1}$ Section for Theoretical and Computational Physics, and
Bergen Computational Physics Laboratory, \\
University of Bergen, 5007 Bergen, Norway \\
\vspace{0.2cm}
$^{2}$ MTA-KFKI, Research Institute of Particle and Nuclear Physics,\\ 
1525 Budapest 114, Hungary\\
\end{center}
}
\normalsize
\date{\today}


\begin{abstract}
The kinetic Freeze Out process of a pion gas  
through a finite layer with time-like normal is considered. 
The pion gas is described by a Boltzmann gas with 
elastic collisions among the pions. 
Within this model, the impact of the in-medium pion mass 
modification on the Freeze Out process is studied. A marginal change 
of the Freeze Out variables temperature and flow velocity and an 
insignificant modification of the frozen out particle distribution function 
has been found. 
\end{abstract}

\section{Introduction}\label{intro}

One of the greatest discoveries in ultra-relativistic heavy-ion physics 
has been the creation of the Quark Gluon Plasma (QGP) at 
Conseil Europeen pour la Recherche Nucleaire (CERN) in 2000 \cite{QGP0} and at 
Relativistic Heavy Ion Collider (RHIC) at Brookhaven National Laboratory (BNL) 
in 2005 \cite{QGP}. In fact, there are compelling experimental signatures which 
transformed the QGP from a theoretical prediction into a precise observational 
science: elliptic flow $v_2$, Jet quenching, Strangeness enhancement 
and Constituent Quark Number scaling 
can hardly be understood without the clear statement that this new state 
of matter has been achieved. However, during the last years it turned out 
that the produced new state of matter is more similar to a strongly 
coupled or strongly interacting Quark Gluon Plasma (sQGP), which 
has more characteristics of a liquid than of a weakly interacting plasma of 
quarks and gluons \cite{sQGP1,sQGP2,sQGP3,sQGP4,sQGP5,sQGP6,sQGP7,sQGP8}; 
for a recent comment on the term "sQGP" see \cite{sQGP9}. While there is 
no doubt about a QGP 
phase transition of Quantum Chromodynamics (QCD) at 
$T_{\rm c} = (173\,\pm 8\,) {\rm MeV}$ \cite{Karsch}, it seems that a QGP of 
freely moving partons can be reached only at higher energy densities and 
temperatures beyond the critical temperature $T_{\rm c}$. 
These new insights imply that further experimental signatures are 
certainly needed to understand not only the physical 
features of sQGP, but also how the new experimental facts do coincide with 
the predictions of the fundamental theory of strong interactions. 
More knowledge of this more complicated new state of 
matter are now expected from the intended experiments at the Large Hadron 
Collider (LHC) at CERN starting up very soon. In the following we will 
not distinguish between the terms QGP and sQGP, but want to keep in mind 
that the new state of matter is more complicated than expected from the 
early theoretical predictions.

The evidence of a QGP cannot be proven directly. Instead, we have to trace 
from the observables at the detectors back to this very early stage of the 
heavy-ion collision.
Obviously, the more accurate the description of the subsequent processes 
after forming the QGP is, the more accurate will be the picture and 
the understanding of this new state of matter.
 
One promising theoretical method in this respect is the hydrodynamical 
approach based on the assumption of local thermal equilibrium. According 
to several theoretical studies, 
the produced QGP reaches a local thermal equilibrium very rapidly within 
$(0.3 - 0.5)\,{\rm fm}/c$ for gluons and $(0.5 - 1.0)\,{\rm fm}/c$ for the 
quarks \cite{Th1,Th2,Th3,Th4}. Experimental data indicate a source size of 
less than 10 fm and less than 10 fm/c time extent. This strongly indicates 
a rapid pre-hadronization \cite{CC94,CM95}, which is also supported by the 
recent observation of constituent quark number scaling of collective flow 
data. Especially, when the expanding system 
reaches a temperature $T \le T_{\rm c}$ hadron states of high multiplicity, 
containing mostly pions, {\it e.g.} \cite{Lit_5, Ratio_5}, are formed.
The pre-thermalization of quarks and quark clusters or pre-hadrons results 
in the local thermalization of pions \cite{Th5,Th6}, and even most of the 
low lying (s) hadronic states.

Subsequently after or even during the hadronization, the chemical and thermal 
Freeze Out (FO) of the hadrons happens, where the hydrodynamical description 
breaks down and transport theoretical approaches are needed. First, the 
inelastic collisions among the hadrons cease, that is 
the so-called chemical FO at $T_{\rm ch}$. Immediately or simultaneously 
followed by the thermal FO at $T_{\rm th}$, where also the elastic collisions 
among the hadrons are abondened. The FO process is essentially the last stage 
of the heavy-ion collision process and the main source for observables. 
An accurate FO description is therefore a basis for an accurate 
understanding of the initial states produced in ultra-relativistic 
heavy-ion collisions. 

A rigorous approach of the FO scenario from first principles  
is given by the Boltzmann Transport Equation which is
a rather difficult assignment of a task. Even more, recently it has
been recognized that the basic assumptions of Boltzmann Transport Equation
are spoiled at the last stages of kinetic FO process \cite{BTE_5,BTE_10},
and a more involved Modified Boltzmann Transport Equation has to be solved.
The reason for that is because the characteristic lenght scale, describing
the change of the distribution function, becomes smaller than the mean
free path $\lambda$ at the last stages of the kinetic FO process.
Thus, phenomenological models which can describe the kinetic FO
process in a simplified manner by taking into account the main features of
a typical FO process only, become rather important. 

Such a phenomenological description of kinetic FO process is usually 
modeled by two different, in some sense even opposite, methods: A FO 
modeling through a hypersurface of zero thickness, and a FO modeling 
through an ininite space-time volume. Recently, the kinetic FO through a 
layer with finite thickness has been developed, for the case of space-like in 
\cite{Spacelike_Layer_1} and for the case of time-like normal in 
\cite{Timelike_Layer_1}; see also \cite{Finite_Layer_1, Finite_Layer_2}. 
This phenomenological approach makes a bridge between these two 
extreme FO models mentioned. So far, the impact of in-medium modifications 
of hadrons on the FO process has been considered only in 
Refs.~\cite{Mod1,Mod2,Zschocke_5}. In our investigation we will apply this 
recently developed FO model and consider a kinetic FO scenario through a finite 
time-like layer to study the impact of in-medium pion mass modification 
on the FO process.  

The paper is organized as follows: 
In Section \ref{TD} we give the needed basics of a transport theoretical 
description of a hot and dense pion gas. The kinetic FO process through a 
finite time-like layer is considered in Section \ref{FO}. The finite temperature 
mass modification of pions, embedded in a Boltzmann gas with 
elastic interactions among the pions, is examined in Section \ref{mass_shift}.
In Section \ref{results_discussions} we present the results obtained, 
and in Section \ref{summary} a summary is given.
Throughout the paper we take $c = \hbar = k_B = 1$.

\section{Transport theoretical description of a pion gas}\label{TD}

Consider a system of $N$ not necessarily conserved number of particles 
described by the one-particle distribution function $f (x,p)$. 
This invariant scalar function is normalized by 
$N = \int d^3 \ve{r}\, d^3 \ve{p} \, f(x,p)$. Throughout the paper we 
consider a dilute pion gas where only elastic scatterings among the pions are allowed, 
so that $x^{\mu} = (t,\ve{r})$ is the four-coordinate 
and $p^{\mu} = (p^0, \ve{p})$ is the four-momentum of the pion with
$p^0 = \sqrt{m_{\pi}^2 + \ve{p}^2}$. Then, the particle four-flow is 
defined by, {\it e.g.} \cite{Laszlo_5},
\bea
N^{\mu} &=& \int \frac{d^3 \ve{p}}{p^0}\,p^{\mu}\,f(x,p)\,,
\label{FO_5}
\eea
and the energy-momentum tensor is
\bea
T^{\mu \nu} &=& \int \frac{d^3 \ve{p}}{p^0}\,p^{\mu}\,p^{\nu}\,f(x,p)\,.
\label{FO_10}
\eea
The four-flow velocity of the medium can be defined as a time-like 
unit tangent vector at the wordline of the particles, {\it i.e.} 
$u^{\mu} = {\rm constant} \times N^{\mu}$ (Eckart's definition). 
However, in case of non-conserved particles like the pions are, 
such a definition would not be convinient. Instead, for non-conserved 
charges or non-conserved particles the four-flow velocity of the medium is 
usually defined as a time-like unit vector parallel to the energy flow, {\it i.e.} 
$u^{\mu} = {\rm constant} \times T^{\mu \nu}\,u_{\nu}$ (Landau's definition):
\bea
u^{\mu}&=&\frac{T^{\mu\nu}\,u_{\nu}}{u_{\rho}\,T^{\rho\sigma}\,u_{\sigma}}\,.
\label{flow_velocity}
\eea
This tensor equation (\ref{flow_velocity}) is by definition valid in any 
frame. 
Obviously, in the rest frame of the gas, RFG, we would have $u^{\mu}_{\rm RFG} 
= (1,0,0,0)$ in Eckart's as well as in Landau's definition of the 
four-flow velocity. 
Any other frame of interest is related to RFG just by a Lorentz boost. 

From {\it eq.}~(\ref{FO_5}) and {\it eq.}~(\ref{FO_10}) one can define three 
linear independent Lorentz invariants: 
scalar particle density $n$, scalar energy density $e$ 
and scalar pressure $P$, given by:
\bea
n &=& N^{\mu} \, u_{\mu}\,,
\label{FO_15}
\\
e &=& u_{\mu} \, T^{\mu \nu} u_{\nu} \,,
\label{FO_20}
\\
P &=& - \frac{1}{3} T^{\mu \nu} \, \Delta_{\mu \nu}\,.
\label{FO_25}
\eea
where $\Delta_{\mu \nu} = g_{\mu \nu} - u_{\mu} u_{\nu}$ projects any
four-vector into the plane orthogonal to $u^{\mu}$; the metric tensor 
$g_{\mu \nu} = {\rm diag} (1,-1,-1,-1)$. We also note the invariant scalar 
entropy density,
\bea
s = S^{\mu}\,u_{\mu} \,, 
\label{entropy_density}
\eea
where the entropy four-current is defined by 
\bea
S^{\mu} &=& - \int \frac{d^3 \ve{p}}{p_0}
\,p^{\mu}\,[f (x, p)\,{\rm ln} f (x, p) - f(x, p)]\,.
\label{entropy_current}
\eea
In our investigation we will consider the FO process of an ultra-relativistic 
heavy-ion collision, and the hot region of the fireball shall be deemed to be 
in chemical ($\mu_{\pi}=0$) and local thermal equilibrium $T(x)$. The 
conserved quantum numbers, ({\it e.g.} baryon number, electric charge, strangeness) 
are zero such that the thermal distribution can 
be characterized by the local temperature parameter $T (x)$ of the fireball. 
Up to temperatures $T \le T_{\rm c}$, most of the particles of such a system are the 
pions \cite{Lit_5, Ratio_5}, which interact via elastic collisions 
with a cross section $\sigma_{\pi \pi}^{\rm elastic}$. 
Thus, we will consider the invariant scalar functions 
{\it eqs.}~(\ref{FO_15}) -- (\ref{entropy_density}) for the case of an  
Boltzmann gas of pions where only elastic scatterings among the 
pions are allowed, and moving with a four-flow velocity 
$u^{\mu}$. The particle distribution function in such a case is homogeneous 
$f (x, p) = f (p)$ and given by the J\"uttner distribution 
\bea
f_{\rm eq} (p,T) &=& g_{\pi} \, \frac{1}{(2 \pi)^3} \, 
{\rm exp} 
\left( \frac{\mu_{\pi} - p^{\mu} u_{\mu}}{T} \right) \,,
\label{Juttner}
\eea
where $g_{\pi} = 3$ is the isospin degeneracy factor.
Since the functions {\it eqs.}~(\ref{FO_15}) -- (\ref{FO_25}) are invariant scalars, 
they can be evaluated in any Lorentz frame. Especially, in the local rest frame
RFG we obtain in case of an pion gas, characterized by the 
distribution function (\ref{Juttner}), the following expressions: 
\bea
n &=& \frac{g_{\pi}}{2\,\pi^2}\,m_{\pi}^2\,T\,K_2 (a)\,,
\label{FO_30}
\\
\nonumber\\
e &=& \frac{g_{\pi}}{8\,\pi^2}\, m_{\pi}^3 \, T \, 
\left[ K_1 (a) + 3\, K_3 (a) \right]\,,
\label{FO_35} 
\\
\nonumber\\
P &=& \frac{g_{\pi}}{2\,\pi^2}\,m_{\pi}^2\,T^2\,K_2 (a)\,,
\label{FO_40} 
\\
\nonumber\\
s &=& \frac{g_{\pi}}{2\,\pi^2}\,m_{\pi}^2\,\left[T\,K_2 (a)
+ \frac{1}{4}\,m_{\pi}\,K_1 (a) 
+ \frac{3}{4}\,m_{\pi}\,K_3 (a) \right]\,,
\label{FO_45}
\eea
where $a = m_{\pi} / T$, and $K_n$ are the Bessel functions of 
second kind, see Appendix A. 
The Equation of State (EoS) $P(n,T)$ of the pion gas follows from 
{\it eq.}~(\ref{FO_30}) and {\it eq.}~(\ref{FO_40}), $P = n\,T$, and the thermodynamical 
relation $T\,s = e + P$ is also satisfied \footnote{Recall, that a 
thermodynamical approach, by means of the canonical potential of an  
pion gas,
\bea
\Omega &=& g_{\pi} \, T \int \frac{d^3 \ve{p}}{(2 \pi)^3} \, {\rm ln}
(1 - {\rm exp} (- p_0 /T)) \,,
\eea
and with the aid of definitions of energy density 
$e = - T^2 \partial \Omega / \partial T$, pressure $P = - \Omega$ and entropy
density $s = \partial \Omega / \partial T$, confirms the findings of 
{\it eq.}~(\ref{FO_35}) - {\it eq.}~(\ref{FO_45}) and the relation $T\,s = e + P$.}. 
In the limit of vanishing  
pion mass $m_{\pi} \rightarrow 0$ we obtain from {\it eq.}~(\ref{FO_35}) and 
{\it eq.}~(\ref{FO_40}) the EoS of an ideal relativistic gas, $e = 3\,P$. 

From these considerations we have seen that in RFG the particle density, 
energy density and pressure are only functions of temperature $T$. 
This implies that in any arbitrary Lorentz frame only 
two thermodynamical unknowns, temperature $T$ and four-flow velocity 
$u_{\mu}$, can enter the problem under consideration. 
To determine both unknowns, we need to have two differential equations,
which can be deduced \footnote{For a proof of {\it eq.}~(\ref{DE_5}) see footnote $3$ 
in \cite{FO_5}, and for {\it eq.}~(\ref{DE_10}) see also the remarks in the Appendix A 
of \cite{Zschocke_5}} from {\it eq.}~(\ref{flow_velocity}) and {\it eq.}~(\ref{FO_20}) 
\cite{FO_5}:
\bea
d u_{\mu} &=& \frac{\Delta_{\mu \nu} \, d T^{\nu \sigma} u_{\sigma}}
{e + P}\,,
\label{DE_5}
\\
\nonumber\\
d e &=& u_{\mu} \, d T^{\mu \nu}\, u_{\nu} \,.
\label{DE_10}
\eea
For the left side of {\it eq.}~(\ref{DE_10}) we obtain from the scalar invariant 
(\ref{FO_35}) the following expression:
\bea
d e &=& \frac{g_{\pi}}{8 \, \pi^2}\, m_{\pi}^3 \, 
\left[ 4 \,a \, K_0 (a) + 8 \,K_1 (a) + 12 \,K_3 (a) \right]\,d T\,.
\label{DE_7}
\eea
These equations (\ref{DE_5}) -- (\ref{DE_7}) can be used to determine the 
temperature and four-flow velocity, while pressure $P$, particle density $n$ 
and entropy density $s$ would follow from their definitions in 
{\it eqs.}~(\ref{FO_15}), (\ref{FO_25}) and (\ref{entropy_density}), respectively. 
The differential of 
the energy-momentum tensor needed in {\it eq.}~(\ref{DE_5}) and {\it eq.}~(\ref{DE_10}) 
follows from {\it eq.}~(\ref{FO_10}),
\bea
d T^{\mu \nu} &=& \int \frac{d^3 \ve{p}}{p^0}\,p^{\mu}\,p^{\nu}\,d f(x,p)\,,
\label{DE_15}
\eea
according to which we still need a differential equation for the one-particle 
distribution function. This will be subject of the next Section.

\section{Freeze Out Process within a finite time-like layer}\label{FO}

The scheme of an ultra-relativistic heavy-ion collision can be subdivided into 
three main stages characterized by their typical temperature parameter $T$: 
First, the initial stage at $T_{\rm c} < T$ where a hot and dense parton gas 
is produced. Second, the stage at $T_{\rm pre-FO} \le T \le T_{\rm c}$ 
where hadrons are formed. And third, the Freeze Out process at temperatures 
$T_{\rm post-FO} \le T \le T_{\rm pre-FO}$ where the hadrons freeze out.
After the complete FO of the hadrons at $T = T_{\rm post-FO}$ the particle 
interactions cease, {\it i.e.} the momentum distribution of the particles is frozen out 
and the hadrons move freely towards the detector.

In this Section we are concerned with the third stage of the collision 
scheme, {\it i.e.} we start our investigation of FO process from the time of collision 
where the expanding system reaches a temperature $T = T_{\rm pre-FO}$ 
and the hadronization of the primary parton gas is considered to be completed.
In the past, the FO process has been usually simulated in two extreme 
scenarios: a sudden FO on a hypersurface with zero proper thickness $L = 0$, 
or a gradual FO process during an infinite time and through an infinite 
space $L\rightarrow \infty$.
 
In this Section we present a model for a gradual FO through a finite 
layer, where the thickness $L$ can be varied from zero to infinity, thus 
making a bridge between the two extreme schemes mentioned above. 
Within such a model 
the FO layer is bounded by two hyper-surfaces: the pre-FO hyper-surface with 
$T = T_{\rm pre-FO}$ where the hydrodynamical description ends, 
and a post-FO hyper-surface with $T = T_{\rm post-FO}$ where all the matter is frozen out. 
A covariant model of kinetic FO process within a finite layer has been recently 
developed, both for the case of space-like \cite{Spacelike_Layer_1} and 
time-like \cite{Timelike_Layer_1} layer; see also 
\cite{Finite_Layer_1, Finite_Layer_2}. 

In order to get an idea about the physical scales of the total FO time $L$ of 
the FO layer, we recall that the collision time $\tau_{\rm coll}$ between the 
pions, which are the dominant hadrons, depends on temperature 
$\tau_{\rm coll} (T) = 12 \, f_{\pi}^4/T^5$ 
\cite{Lit_10,Lit_15,Lit_20}, where $f_{\pi} = 92.4 \,{\rm MeV}$ is the pion 
decay constant ($1 = 0.19733\,{\rm GeV}\, {\rm fm}$). 
At the pre-FO side of the layer there is a temperature of 
$T_{\rm pre-FO} \approx 175 \, {\rm MeV}$ and the collision time is small: 
$\tau_{\rm coll} \simeq 1.1\,{\rm fm}/c$. 
The proper thickness $L$ in time of the FO layer is taken typically of the
order of a few (at least one) collision time at $T = T_{\rm pre-FO} \sim T_{\rm c}$, 
{\it i.e.} $L \sim (5 ... 10) {\rm fm}/c\,$.

Here, we will not repeat the theoretical developments of 
\cite{Spacelike_Layer_1,Timelike_Layer_1,Finite_Layer_1, Finite_Layer_2} 
in detail, but should consider the essential steps relevant for our 
investigations. 

To describe the gradual FO process, the one-particle distribution function 
is decomposed into two components, an interacting part $f_i$ and 
a frozen-out part $f_f$,
\bea
f (x,p) &=& f_i (x,p) + f_f (x,p) \,.
\label{FO_60}
\eea
During the FO process the number of 
interacting particles decreases from pre-FO to post-FO side, where by 
definition the number of interacting particles tends to zero. 
As boundary conditions we assume on the pre-FO side 
of the layer a thermal equilibrium, {\it i.e.} a J\"uttner distribution 
(\ref{Juttner}) for $f_i$ and $f_f=0$, while on the post-FO side $f_i$ 
vanishes; for an illustration see also {\it fig.}~1 in \cite{Spacelike_Layer_1}.

The space-time evolution of the interacting and non-interacting components
during the FO should be modeled by the Boltzmann Transport Equation (BTE). 
We will use the relaxation time approximation and apply the escape rate
$P_{esc}(x,p)$, describing the escape of particles
from the interacting component $f_i$ to the non-interacting component $f_f$.
The FO is a strongly directed process, {\it i.e.} the gradient in one preferred 
FO direction, $d \sigma_{\mu}=(d \sigma_0, d \ve{\sigma})$, 
is much stronger than the changes in the 
perpendicular 
directions, we can neglect these FO gradients in the perpendicular directions.
Then, the BTE can be transformed into the following differential equations
\cite{Spacelike_Layer_1,Timelike_Layer_1,Finite_Layer_1, Finite_Layer_2}:
\bea
d \sigma^{\mu} \, \partial_{\mu} f_i (x,p) &=& - P_{esc}(x,p)\,f_i (x,p)
 + \; \frac{1}{\tau_{\rm th}} [f_{\rm eq} (p) - f_i (x,p)]\,,
\label{FO_equation_5}
\\
d \sigma^{\mu} \, \partial_{\mu} f_f (x,p) &=& P_{esc}(x,p)\,f_i (x,p)\,.
\label{FO_equation_10}
\eea
Troughout the paper we shall use the notation  
\bea
\partial_{\mu} \, f (x,p) &\equiv& \frac{\partial}{\partial x^{\mu}} \,f(x,p)
\,,
\label{FO_equation_11}
\eea
{\it i.e.} expressions like (\ref{FO_equation_11}) are not infinitesimal 
quantities.  Note, the finite normal vector on hypersurface is normalized 
by $d \sigma_{\mu}\;d \sigma^{\mu} = \pm 1$ where upper sign 
is for time-like and lower sign is for space-like normal, respectively; 
note that $d \sigma_{\mu}$ is also not an infinitesimal quantity but a 
finite vector (an explicit expression for time-like normal is given below). 
Here, $p_{\mu} = (p_0, \ve{p})$ is the four-momentum of particle, and 
$x_{\mu} = (t, \ve{r})$ is the four-coordinate of particle.  
The second term in {\it eq.}~(\ref{FO_equation_5}) 
is the re-thermalization term (see below), which describes how fast the 
system relaxes into some thermalized distribution function $f_{\rm eq}$ 
during a characteristic time scale $\tau_{\rm th}$. 

A Lorentz invariant expression for the escape rate is given by 
\cite{BTE_10,Spacelike_Layer_1,Timelike_Layer_1,Finite_Layer_1,Finite_Layer_2}. 
\bea
P_{esc} (x,p) &=& \frac{1}{\tau_0} \frac{L}{L - x^{\mu}\,d \sigma_{\mu}} \, 
\frac{p^{\mu}\,d \sigma_{\mu}}{p^{\mu}\,u_{\mu}} \,
{\Theta} (p^{\mu} d \sigma_{\mu})\,,
\label{FO_equation_15}
\eea
where $\tau_0$ is the characteristic FO time. 
The $\Theta-{\rm function}$ is the Bugaev cut-off factor \cite{Bugaev}, which
is important only for the FO in space-like direction.

We have to insert the escape rate $P_{esc}$ into {\it eqs.}~(\ref{FO_equation_5}) 
and (\ref{FO_equation_10}). In the following we will consider the FO process 
of a pion gas through a finite layer with a time-like normal 
$d \sigma_{\mu}\,d \sigma^{\mu}=+1$, so that $L$ becomes a thickness in time.
We will work in the Rest Frame of the FO front, RFF, where $d \sigma_{\mu} = 
(1,0,0,0)$. Thus, we obtain 
the following set of differential equations:
\bea
\partial_t \, f_i &=& - \frac{1}{\tau_0}\,\left(
\frac{L}{L - t}\right)
\left(\frac{p^0}{p_{\mu} \, u^{\mu}} \right) f_i
\; + \; \frac{1}{\tau_{\rm th}} [f_{\rm eq} (p) - f_i]\,,
\label{FO_65}
\\
\nonumber\\
\partial_t \, f_f &=& + \frac{1}{\tau_0}\,
\left(\frac{L}{L - t}\right)
\left(\frac{p^0}{p_{\mu} \, u^{\mu}} \right) f_i\,.
\label{FO_70}
\eea
Note again, the first term in {\it eqs.}~(\ref{FO_65}) and (\ref{FO_70}) describes 
the transition of the pions from the interacting to the frozen out component. 
The second term in {\it eq.}~(\ref{FO_65}) is the 
re-thermalization term \cite{FO_5, FO_10} which describes how the interacting 
component relaxes to some thermal distribution $f_{\rm eq}$, where the 
parameters of it, $T(t), u_{\mu}(t)$, have to be calculated from the 
conservation laws. The strength of both terms are 
characterized by their typical time scales, the characteristic Freeze Out time 
$\tau_0$ and relaxation time $\tau_{\rm th}$, respectively 
\cite{Timelike_Layer_1,FO_5,FO_10,tau_1,FO_15,tau_2}. 

In the case of fast re-thermalization $\tau_{\rm th} \ll \tau_0$, 
the interacting component can be choosen as equilibrated J\"uttner distribution 
for all the times \cite{Spacelike_Layer_1}. Then we obtain, with the aid of 
{\it eq.}~(\ref{DE_15}), for the energy-momentum tensor of interacting component
\bea
\frac{d T^{\mu \nu}_i}{d t} &=& 
\int \frac{d^3 \ve{p}}{p^0}\,p^{\mu}\,p^{\nu}\, \partial_t \, f_i (x,p) 
\nonumber\\
&=&  - \frac{1}{\tau_0}\,\left(\frac{L}{L - t}\right) \int 
\frac{d^3 \ve{p}}{p^0}\,p^{\mu}\,p^{\nu}\, \left(\frac{p^{0}}{p_{\mu} 
\, u^{\mu}} \right) f_{\rm eq} (p, T(t), u_{\mu}(t))\,.
\label{FO_75}
\eea
In the following we will give the components of {\it eq.}~(\ref{FO_75}) in the RFF:
\bea
\frac{d T^{0 0}_i (t, v, T, m_{\pi})}{d t} &=& \frac{1}{\tau_0}
\frac{L}{L - t} \frac{n T}{4} \frac{1}{\gamma \, v}
\left(G^{-}_2 (m_{\pi}, v, T) - G^{+}_2 (m_{\pi}, v, T) \right)\,,
\label{freezeout_15}
\\
\nonumber\\
\frac{d T^{0 x}_i (t, v, T, m_{\pi})}{d t} &=& \frac{1}{v}
\frac{d T^{0 0}_i (t, v, T, m_{\pi})}{d t}
\nonumber\\
\nonumber\\
&& +  
\frac{1}{\tau_0} \frac{L}{L - t}\frac{n T}{2} \frac{b^2}{\gamma\,v} 
\left( (3 + v^2) K_2 (a) + a \, K_1 (a) \right)\,,
\label{freezeout_20}
\\
\nonumber\\
\frac{d T^{x x}_i (t, v, T, m_{\pi})}{d t} &=& \frac{1}{v}
\frac{d T^{0 x}_i (t, v, T, m_{\pi})}{d t}
\nonumber\\
\nonumber\\
&& - \frac{T}{\gamma\,v}  \left(\frac{d N^{x}_i (t, v, T, m_{\pi})}{d t}
- \frac{1}{v} \frac{d N^{0}_i (t, v, T, m_{\pi})}{d t} \right)
\nonumber\\
\nonumber\\
&& + \frac{1}{\tau_0} \frac{L}{L - t}
\frac{n T}{2} a \, b
\left( \frac{1}{v^2} (1 + 3 v^2) K_2 (a) + b \, K_1 (a) \right)\,,
\label{freezeout_25}
\eea
and the needed components of time derivative of 
the particle four-current are given by 
\bea
\frac{d N^{0}_i (t, v, T, m_{\pi})}{d t} &=& \frac{1}{\tau_0}
\frac{L}{L - t} \frac{n}{4} \left(
G^{-}_1 (m_{\pi}, v, T) - G^{+}_1 (m_{\pi}, v, T) \right)  \,,
\label{freezeout_5}
\\
\nonumber\\
\frac{d N^{x}_i (t, v, T, m_{\pi})}{d t} &=& \frac{1}{v}
\frac{d N^{0} (t, v, T, m_{\pi})}{d t} + \frac{1}{\tau_0} 
\frac{L}{L - t}
\frac{n}{4} \left( \frac{4 a K_1 (a) }{v} + \frac{2 a^2 K_0 (a)}{v} \right)\,,
\nonumber\\
\label{freezeout_10}
\eea
were we recall that $a = m_{\pi}/T$ and $b = \gamma\,a$; 
the functions $G_n^{\pm}$ and $K_n$ are defined in the Appendix A. 
According to these expressions we also need the invariant 
scalar pion density $n$ defined in {\it eq.}~(\ref{FO_15}), and according to 
{\it eq.}~(\ref{DE_5}) we also need the pressure $P$ defined in {\it eq.}~(\ref{FO_25}). 
Since they are invariant scalars, we can take the explicit expressions given 
in {\it eq.}~(\ref{FO_30}) and {\it eq.}~(\ref{FO_40}) evaluated in the RFG for a J\"uttner 
distribution. By inserting these results and {\it eqs.}~(\ref{freezeout_15}) -- 
(\ref{freezeout_25}) into {\it eq.}~(\ref{DE_5}) and {\it eq.}~(\ref{DE_10}), we obtain a 
set of two differential equations for the two unknowns $T$ and $v$. 
Taking $u^{\mu}_{\rm RFF} = \gamma (1,v,0,0)$, 
$u_{\mu}^{\rm RFF} = \gamma (1,-v,0,0)$, 
$d u_0^{\rm RFF} = \gamma^3\,v\,d v$ and $d u_x^{\rm RFF} = - \gamma^3\,d v$, 
we obtain explicitly in RFF:
\bea
\frac{d T}{d t} &=& \frac{8\,\pi^2}{g_{\pi}\,m_{\pi}^3}
\left(4 \, a \,K_0 (a)
+ 8 K_1 (a) + 12 K_3 (a) \right)^{- 1}\,\gamma^2\, 
\left(\frac{d T^{0 0}_i}{d t} - 2 v \frac{d T^{0 x}_i}{d t} + v^2 
\frac{d T^{x x}_i}{d t} \right) ,
\label{DE_20}
\\
\nonumber\\
\frac{d v}{d t} &=& \frac{2 \pi^2}{g_{\pi}\, m_{\pi}^3 \,T} 
\left(\frac{1}{4} K_1 (a)
+ \frac{3}{4} K_3 (a) + \frac{1}{a}  K_2 (a) \right)^{- 1} 
\left( - v \frac{d T^{0 0}_i}{d t} 
+ (1 + v^2) \frac{d T^{0 x}_i}{d t} - v \frac{d T^{x x}_i}{d t}\right).
\nonumber\\
\label{DE_25}
\eea
Notice, that the degeneracy factor $g_{\pi}$ is actually 
cancelled against the same factor contained in the scalar particle density 
$n$ of energy-momentum components, see {\it eq.}~(\ref{FO_30}).
In the limit of vanishing pion mass $m_{\pi} \rightarrow 0$ the 
{\it eqs.}~(\ref{DE_20}) and (\ref{DE_25}) simplify to 
\bea
\frac{d T}{d t} &=& - \frac{1}{\tau_0} \, \frac{L}{L - t} \, 
\frac{1}{4}\,T\,\gamma\,,
\nonumber\\
\label{DE_26}
\\
\frac{d v}{d t} &=&  - \frac{1}{\tau_0} \, \frac{L}{L - t} \,
\frac{1}{4}\,\frac{v}{\gamma}\,,
\label{DE_27}
\eea
in agreement with the corresponding limit given in {\it eq.}~(9) 
in \cite{Timelike_Layer_1} \footnote{Note, in case of a massless pion gas 
we have, due to J\"uttner distribution,
\bea
e &=& g_{\pi}\,\frac{4\,\pi}{(2 \pi)^3} \int\limits_{0}^{\infty} d p\, p^3\, 
{\rm exp}(-p/T) = 3\,g_{\pi}\,T^4/\pi^2\,,
\eea
while in \cite{Timelike_Layer_1} a Bose gas was assumed for the EoS:
\bea
e_{\rm Bose} &=& g_{\pi}\,\frac{4\,\pi}{(2 \pi)^3} 
\int\limits_{0}^{\infty} d p\, p^3\,
\left[{\rm exp}(p/T) - 1\right]^{- 1} = g_{\pi}\,\pi^2\,T^4 / 30\,.
\eea
The difference is only marginal. However, in {\it eq.}~(9) of 
Ref.~\cite{Timelike_Layer_1} we have actually to imply the relation 
$n_{\rm Bose} = e_{\rm Bose}/(3\,T) = g_{\pi} \, \pi^2\,T^3 / 90$ 
valid for massless pions, to get the agreement stated above.}.
The system of two differential equations (\ref{DE_20}), (\ref{DE_25}), 
together with the invariant scalars $n, e, P$ in {\it eqs.}~(\ref{FO_30}) -- 
 (\ref{FO_40}) and the components of energy-momentum tensor 
given in {\it eqs.}~(\ref{freezeout_15}) -- (\ref{freezeout_25}) 
constitute a closed set of equations for the unknowns $T (t)$ and $v (t)$ 
inside the FO layer. Once these both unknows have been determined self 
consistently, all other quantities like particle density $n(t)$, pressure 
$P(t)$ and entropy density $s(t)$ inside the finite layer can be determined 
by their expressions given in {\it eqs.}~(\ref{FO_30}) -- (\ref{FO_45}).

In our investigation, we are interested on the impact of in-medium pion 
mass modification on the FO process. Therefore, we have to implement a 
temperature dependent pion mass $m_{\pi} (T)$ in the given equations. 
The results of such an investigation can then be compared with the 
corresponding findings where a vacuum pion mass $m_{\pi}$ or massless 
pions are implemented. We note, however, that in-medium modified pion mass 
not only changes slightly the distribution function, but also the 
pionic EoS and correspondingly the expansion dynamics, {\it e.g.} \cite{Wambach}. 
We note, that such a consideration could also be investigated, 
within the expanding model very recently developed in \cite{V_Magas1}. 

\section{Pion Mass Shift at finite temperature}\label{mass_shift}

Typical conditions inside the FO layer are high temperatures, 
typically between $100\,{\rm MeV} \le T \le 175\,{\rm MeV}$. Such  
extreme conditions imply a 
strong modifications of the hadrons in respect to their mass, coupling 
constants and decay rates are expected. One aim of our investigation
is to evaluate how strong the impact of pion mass shift on kinetic FO process 
of a pion gas is. The kinetic FO process of a pion gas concerns elastic interactions among these 
particles and how these elastic scatterings cease. Therefore, we have to determine 
the mass modification of pions embedded in a pion gas with elastic scatterings  
among these particles. However, the impact of non-elastic interactions on mass modification 
of pions becomes relevant for a description of the chemical FO process. 

\subsection{Pion mass in vacuum}

First, let us briefly re-consider the mass of a pion in vacuum, 
defined as pole mass of the pion propagator:
\bea
\Pi_{\pi}^a (p) &=& i \, \int d^4 x \; {\rm e}^{i p x} \;
\langle {\rm T}_{\rm W} \,\hat{\Phi}^a (x) \hat{\Phi}^{a\;\dagger} (0) \rangle_0
\nonumber\\
&=& \frac{1}{p^2 - \stackrel{\rm o}{m}_{\pi}^2 - \Sigma_{\pi}^a (p)
+ i \epsilon}\;,
\label{massshift_5}
\eea
where $\langle \hat{\cal O} \rangle_0 = \langle 0 | {\cal O} | 0 \rangle$ is the vacuum 
expectation value of an operator $\hat{\cal O}$, $a = 1,2,3$ is the isospin index, ${\rm T}_{\rm W}$ 
is the Wick time ordering, $\hat{\Phi}^a$ is the second-quantized pion field operator 
and $\Sigma_{\pi}^a$ is the self energy of pion $a$ in vacuum. 
The parameter $\stackrel{\rm o}{m}_{\pi}$ is the so-called bare pion mass 
which enters the Lagrangian of the effective hadron model of QCD. 
The physical pion mass is defined as the pole of 
propagator {\it eq.}~(\ref{massshift_5}), {\it i.e.} as the selfconsistent solution of
\bea
m_{\pi}^2 &=& \stackrel{\rm o}{m}_{\pi}^2 
\,+ \; {\rm Re} \, \Sigma_{\pi} (p^2 = m_{\pi}^2)\,.
\label{massshift_10}
\eea
The vacuum pion mass is $m_{\pi}^0 = 135.04\,{\rm MeV}$, 
$m_{\pi}^{\pm} = 139.63\,{\rm MeV}$.
Despite enormous effort in Quantum field theory, a rigorous derivation of 
vacuum pion mass from first principles of QCD has not been found so far. 
That means the vacuum mass of any hadron 
can not been obtained from fundamental QCD without further assumption 
or new parameters. 
However, there are promising and sophisticated approaches which have
provided some insights into this very involved issue. Among them there 
is Chiral Perturbation Theory, Lattice gauge theory, Current 
algebra, Dyson-Schwinger approach, Nambu-Jona-Lasinio model and QCD sum rules, 
which provide a link between the quark degrees of 
freedom of underlaying QCD and the hadronic degrees of freedom. Such 
approaches allow a derivation of the so-called Gell-Mann--Oakes--Renner (GOR) 
relation \cite{GOR_1, GOR_2},
\bea
m_{\pi}^2 \, f_{\pi}^2 
&=& - 2 \, m_q \, \langle \hat{\overline q} \hat{q} \rangle_0 \,,
\label{GOR_5}
\eea
which allows to determine the pion mass in vacuum from the microscopic 
QCD quantities current quark mass $m_q$ and chiral quark condensate at $x=0$,
\bea
\langle \hat{\overline q} \hat{q} \rangle_0 &=& \frac{1}{2} \; 
\langle \hat{\overline u} \hat{u} + \hat{\overline d} \hat{d} \rangle_0  \,.
\label{chiral_condensate}
\eea
Typical values are $m_q = (m_u + m_d)/2 = 5.5 \,{\rm MeV}$,  
$\langle \hat{\overline q} \hat{q} \rangle_0 = - (245\,{\rm MeV})^3$; 
recall, $f_{\pi} = 92.4\,{\rm MeV}$ is the pion decay constant. 
With these given numerical values, the GOR relation yields $m_{\pi} = 138\,{\rm MeV}$.
The pion decay constant can be defined by 
\bea
\langle 0 | \hat{A}^a_{\mu} (0) | \pi^b (q) \rangle &=& - i \, f_{\pi}\, q_{\mu} \,\delta_{a b}\,,
\label{pion_decay_constant_5}
\eea
where the isospin indices $a,b=1,2,3$ and 
\bea
\hat{A}^a_{x} (x) &=& \hat{\overline \Psi} (x) \,\gamma_{\mu} \,\gamma_5 \;\frac{\tau^a}{2}\;\hat{\Psi} (x)
\label{axial_current}
\eea
is the axial vector current operator, {\it e.g.} \cite{Weise}; $\tau^a$ are the Pauli spin matrices, 
and $\hat{\Psi} = (\hat{u} \;,\; \hat{d})^{\rm T}$.
From {\it eq.}~(\ref{pion_decay_constant_5}) we obtain 
the expression for the pion decay constant in vacuum,
\bea
f_{\pi} &=& i\, \frac{q^{\mu}}{q^2}\;\langle 0 | \hat{A}^a_{\mu} (0) | \pi^a (q) \rangle\,,
\label{pion_decay_constant_7}
\eea
where, due to isospin symmetry, $a$ can be chosen arbitrarily, {\it e.g.} $a=3$ 
({\it i.e.} there is no sum over index $a$). It should be noted that with the aid of 
LSZ reduction formalism \cite{LSZ1,LSZ2}, the expression (\ref{pion_decay_constant_7}) 
can be rewritten as vacuum expectation value over the scalar operator
\bea
\hat{f_{\pi}} (0) &\equiv& i\, \frac{q^{\mu}}{q^2}\;
\int d^4 y \; {\rm exp} (- i q y) \;
\left( \Box_{y} + m_{\pi}^2 \right)
{\rm T}_W \hat{A}_{\mu}^a (0) \,\hat{\Phi}^{a\;\dagger} (y)\, \,,
\label{LSZ_10}
\eea
that means we basically have 
\bea
f_{\pi} &=& \langle \hat{f}_{\pi} \rangle_0 \,.
\label{LSZ_5}
\eea

\subsection{Pion mass at finite temperature}

\begin{figure}
\begin{center}
\hspace{-0.5cm}\includegraphics[angle=270,scale=0.5]{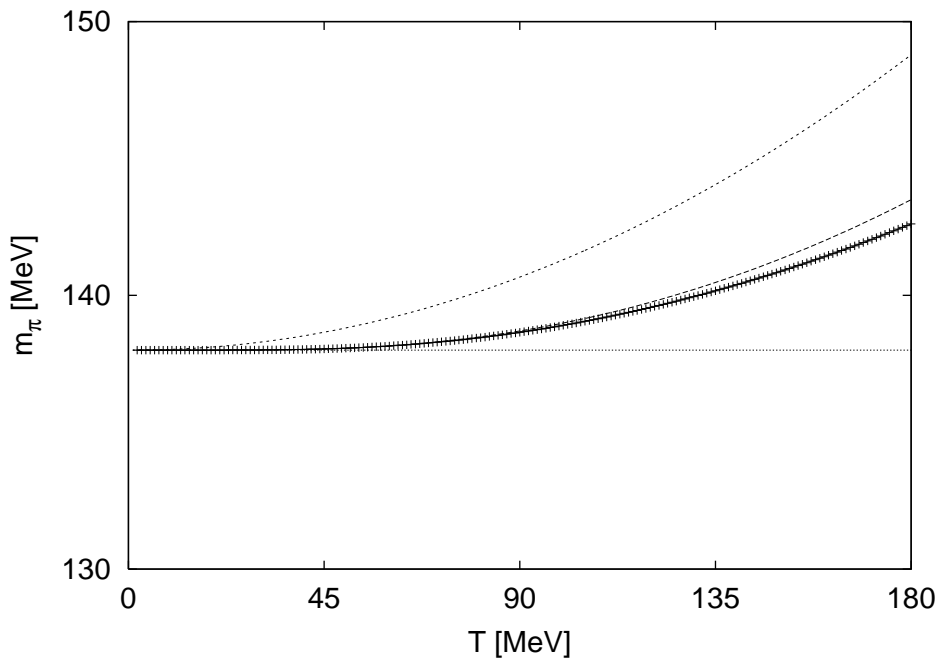}
\caption{The temperature dependent pion mass $m_{\pi} (T)$: Thick solid 
linepoints show the pion mass for a Boltzmann gas according to {\it eq.}~(\ref{pion_mass_shift_T_1}).
The straight dotted line represents a constant pion mass in vacuum,
$m_{\pi} = 138\,{\rm MeV}$, according to the GOR relation
in vacuum (\ref{GOR_5}). The dahed line represents the pion mass shift in case of a 
Bose distribution, {\it i.e.} when in {\it eq.}~(\ref{pion_mass_shift_T_1}) function 
$B_1$ is replaced by function $B_2$. The difference between a pion mass shift of a Boltzmann gas 
and of a Bose gas are marginal in the temperature region considered. 
The dotted line represents the result of ChPT according to {\it eq.}~(\ref{pion_mass_shift_T_2}), 
allowing a comparison of our results with ChPT.}
\label{fig:mass_shift}
\end{center}
\end{figure}

Now let us consider the case of pions in medium.  
At finite temperature the pions move in a hot and dense bath of hadrons. 
Accordingly, the in-medium pion propagator reads 
\bea
\Pi_{\pi}^a (p,T) &=& i \, \int d^4 x \; {\rm e}^{i p x} \;
\langle {\rm T}_{\rm W} \; \hat{\Phi}^a (x) 
\hat{\Phi}^{a\;\dagger} (0) \rangle_{\rm T}
\nonumber\\
&=& \frac{1}{p^2 - \stackrel{\rm o}{m}_{\pi}^2 - \Sigma_{\pi}^a (p,T)
+ i \epsilon}\;.
\label{massshift_15}
\eea
The thermal Gibbs average $\langle \hat{\cal O} \rangle_{\rm T}$ of an operator $\hat{\cal O}$ 
is defined by 
\bea
\langle \hat{\cal O} \rangle_{\rm T} &=& 
\frac{{\rm Tr}\;\langle \hat{O} \;{\rm exp} (- \beta \,\hat{H})\rangle}
{{\rm Tr}\; \langle {\rm exp} (- \beta \,\hat{H})\rangle}
= \frac{\sum\limits_{n=0}^{\infty} \; \langle n| \hat{O} \;{\rm exp} (- \beta \,\hat{H}) | n \rangle}
{\sum\limits_{n=0}^{\infty} \; \langle n | {\rm exp} (- \beta \,\hat{H})| n \rangle}\;,
\label{Gibbs_5}
\eea
where $\beta = 1/T$. In general, for temperatures $T \ge T_{\rm c}$ quark and gluon degrees of freedom 
would have to be included. i.e. the sum in {\it eq.}~(\ref{Gibbs_5}) runs in general over quark and gluon 
eigenstates and over hadron eigenstates of $\hat{H}$. However, in our investigation the temperatures 
are $T\le T_{\rm c}$, {\it i.e.} the operator $\hat{H}$ is here an effective Hamiltonian describing the hadron 
system under consideration. Accordingly, we consider $|n\rangle$ as hadron eigenstates of $\hat{H}$, and 
the sum runs over the spectrum of all these hadron eigenstates $|n\rangle$. Furthermore, throughout our considerations 
we are interested on Gibbs average of operators at $x=0$, 
{\it i.e.} we always have $\langle \hat{\cal O} \rangle_{\rm T} = \langle \hat{\cal O} (x=0) \rangle_{\rm T}$.
As in vacuum, the pion pole mass at finite temperature is defined as selfconsistent solution of 
\bea
m_{\pi}^2 (T) &=& \stackrel{\rm o}{m}_{\pi}^2
\,+\; {\rm Re} \, \Sigma_{\pi} (p^2 = m_{\pi}^2 (T), T)\,.
\label{massshift_20}
\eea
From this equation it becomes obvious why there is an in-medium modification
of pion mass: just because the self energy $\Sigma_{\pi} (p,T)$ is now 
temperature dependent which changes the position of pole mass compared to 
vacuum. Such a change of the pole position is caused by elastic interactions and non-elastic 
interactions among the particles. To find an expression for 
the temperature dependence of the pion mass, we will follow a similar way as 
in vacuum, {\it i.e.} we will apply the in-medium GOR relation. 
Especialy, it is a well-known fact, that the GOR-relation continues to 
be valid also at finite temperatures, at least up to order ${\cal O} (T^6)$
\cite{GOR_T_1,GOR_T_2,GOR_T_3,GOR_T_4,GOR_T_5,Dyson_Schwinger_1,
Dyson_Schwinger_2},
\bea
m_{\pi}^2 (T) \, f_{\pi}^2 (T) 
&=& - 2 \, m_q \, \langle \hat{\overline q} \hat{q} \rangle_T \,.
\label{GOR_10}
\eea
We will take relation (\ref{GOR_10}) as a given fact, 
which would allow a determination of the temperature dependence of pion mass;  
the current quark mass $m_q$ as a fundamental parameter of QCD is of 
course independent of temperature, while the constituent quark mass $M_q$ is 
temperature dependent. For that we would have to know the temperature dependence of 
the chiral condensate and of the pion decay constant. That means, in generalization of the vacuum  
expectation values in {\it eqs.}~(\ref{chiral_condensate}) and (\ref{LSZ_5}), we have now to determine the expressions 
\bea
\langle \hat{\overline q} \hat{q} \rangle_T &=&
\frac{\sum\limits_{n=0}^{\infty} \; \langle n| \hat{\overline q} \hat{q} \;{\rm exp} (- \beta \,\hat{H}) | n \rangle}
{\sum\limits_{n=0}^{\infty} \; \langle n | {\rm exp} (- \beta \,\hat{H})| n \rangle}\,,
\label{expression1}
\\
\nonumber\\
f_{\pi} (T) &=& 
\frac{\sum\limits_{n=0}^{\infty} \; \langle n| \hat{f_{\pi}} \;{\rm exp} (- \beta \,\hat{H}) | n \rangle}
{\sum\limits_{n=0}^{\infty} \; \langle n | {\rm exp} (- \beta \,\hat{H})| n \rangle}\,,
\label{expression2}
\eea
where the operator $ \hat{f_{\pi}}$ is defined in {\it eq.}~(\ref{LSZ_10}). Both of these expressions, 
$\langle \hat{\overline q} \hat{q} \rangle_T$ and $f_{\pi} (T) = \langle \hat{f}_{\pi}\rangle_T$, 
have been evaluated by means of several approaches. The aim is to find an expression 
both for chiral condensate and pion decay constant at finite temperature and consistently 
within the kinetic model description. Especially, in our approach 
there are only pions. Accordingly, the Gibbs average of an operator $\hat{\cal O}$ runs over the 
diagonal pion states only, {\it i.e.} $\langle \pi | \hat{\cal O} | \pi \rangle$, 
$\langle \pi \pi | \hat{\cal O} | \pi \pi \rangle, ... $ \cite{Ioffe1,Ioffe2}, but does not include the diagonal 
matrix elements of heavier hadrons like Kaons, $\langle K^{\pm} | \hat{\cal O} | K^{\pm} \rangle$, etc. 
Such an approximation can be justified, since at low temperatures $T \le T_{\rm c}$, pion states dominate 
the thermal average, while states containing heavier mesons with a mass $m_n > m_{\pi}$ are weighted 
with their corresponding Boltzmann factor $\sim {\rm exp} (- m_n/T)$, {\it i.e.} they are exponentially suppressed 
\cite{Lit_5, Ratio_5}.
For instance, in Ref. \cite{Lit_5,F_pi_5} contributions of heavier mesons to the chiral condensate
at finite temperature have been studied, where only marginal corrections were found:
$5$ percent corrections at $T = 100\,{\rm MeV}$ and $10$ percent corrections at $T = 150\,{\rm MeV}$.
The Gibbs average (\ref{Gibbs_5}) can be further approximated by one-pion states \cite{F_pi_5,Bochkarev,Ioffe3},
\bea
\langle \hat{\cal O} \rangle_{\rm T} &=&
\langle \hat{\cal O} \rangle_0 + \sum \limits_{n=1}^{3}\;\int \frac{d^3 p}{(2 \pi)^3\;2\,p^0}\;
\langle \pi^n (p) | \hat{O} | \pi^n (p) \rangle\;{\rm exp} \left(- \frac{p^0}{T} \right) 
+ {\cal O} \left(T^4 \right)\;.
\label{Gibbs_10}
\eea 
The contributions of next higher order ${\cal O} (T^4)$ are considered in Appendix C,
according to which we will neglect multi-pion states in the temperature region considered, see also 
Refs.~\cite{Ioffe1,Ioffe2}. The pion states are normalized by 
\bea
\langle \pi^n (p_1) | \pi^m (p_2) \rangle &=& 
2 \;{p_1^0} \;(2 \pi)^3 \;\delta^{(3)} (\ve{p}_1 - \ve{p}_2)\; \delta_{n m} \,,
\eea
with isospin indices $n,m=1,2,3$. With the aid of {\it eq.}~(\ref{Gibbs_10}) and according to {\it eqs.}~(\ref{expression1}) 
and (\ref{expression2}), we obtain 
\bea
\langle \hat{\overline q} \hat{q} \rangle_{\rm T} &=&
\langle \hat{\overline q} \hat{q} \rangle_0 + \sum \limits_{n=1}^{3}\;\int \frac{d^3 p}{(2 \pi)^3\;2\,p^0}\;
\langle \pi^n (p) | \hat{\overline q} \hat{q} | \pi^n (p) \rangle\;{\rm exp} \left(- \frac{p^0}{T} \right)
+ {\cal O} \left(T^4 \right)\,,
\label{Gibbs_15}
\\
\nonumber\\
f_{\pi} (T) &=& f_{\pi} + i \frac{q^{\mu}}{q^2} 
\sum \limits_{n=1}^{3}\;\int \frac{d^3 p}{(2 \pi)^3\;2\,p^0}\;
\langle \pi^n (p) | \hat{A}_{\mu}^a | \pi^a (q) \pi^n (p) \rangle
\;{\rm exp} \left(- \frac{p^0}{T} \right) + {\cal O} \left(T^4 \right)\,,
\nonumber\\
\label{Gibbs_20}
\eea
where in the last line the LSZ reduction of pion state $\pi^a (q)$ has been transformed back 
just after the thermal Gibbs average.
The pion matrix elements can be determined using soft-pion theorem \cite{F_pi_5,soft1,soft2,soft3,Toki} 
by means of which we obtain (see {\it eqs.}~(\ref{matrixelement_1}) - (\ref{matrixelement_4}) in Appendix B)
\bea
\langle \hat{\overline q} \hat{q} \rangle_T &=& \langle \hat{\overline q} \hat{q} \rangle_0
\left( 1 - \frac{1}{8} \frac{T^2}{f_{\pi}^2}\; B_1\left(\frac{m_{\pi}}{T}\right) \right) 
+ {\cal O} \left(T^4 \right)\,,
\label{expression3}
\\
\nonumber\\
f_{\pi} (T) &=& f_{\pi}
\left( 1 - \frac{1}{12} \frac{T^2}{f_{\pi}^2} \; B_1\left(\frac{m_{\pi}}{T}\right) \right) 
+ {\cal O} \left(T^4 \right)\,,
\label{expression4}
\\
\nonumber\\
m_{\pi} (T) &=& m_{\pi} \left( 1 + \frac{1}{48}\,\frac{T^2}{f_{\pi}^2} \;
B_1 \left(\frac{m_{\pi}}{T}\right) \right) + {\cal O} \left(T^4 \right)\,,
\label{pion_mass_shift_T_1}
\eea
for the next higher order ${\cal O} (T^4)$ see Appendix C.
In order to derive relation (\ref{pion_mass_shift_T_1}), we have inserted 
{\it eqs.}~(\ref{expression3}) and (\ref{expression4}) into GOR relation at finite temperature 
{\it eq.}~(\ref{GOR_10}), as well as the GOR in vacuum {\it eq.}~(\ref{GOR_5}) has been applied. The function is
\footnote{In case we would have taken a Bose distribution in {\it eq.}~(\ref{Gibbs_10}),
then in {\it eqs.}~(\ref{expression3}) - (\ref{pion_mass_shift_T_1}) the function $B_1$ 
would have to be replaced by the function $B_2$:
\bea
B_2 (z) &=&
\frac{6}{\pi^2}\int\limits_{z}^{\infty} dx \;\sqrt{x^2 - z^2}\;\frac{1}{{\rm exp}(x) - 1}\;,\quad
\lim_{z \rightarrow 0} B_2 (z) = 1\;.
\label{B_2}
\eea
However, the difference for pion mass shift using $B_1$ or $B_2$ is marginal in the temperature region considered,
see {\it fig.}~\ref{fig:mass_shift}.} 
\bea
B_1 (z) &=& \frac{6}{\pi^2}\int\limits_{z}^{\infty} dx \;\sqrt{x^2 - z^2}\;{\rm exp}(- x) 
= \frac{6}{\pi^2}\; z \;K_1 (z) \;,\quad
\lim_{z \rightarrow 0} B_1 (z) =  \frac{6}{\pi^2} \;.
\label{function_B1}
\eea
The results (\ref{expression3}) - (\ref{pion_mass_shift_T_1}) are consistently valid for a pion gas 
approximated by a Boltzmann gas, {\it i.e.} with elastic interactions among the pions.
Here, we will require the applicability of {\it eqs.}~(\ref{expression3}) - (\ref{pion_mass_shift_T_1}) 
in a temperature region $T \le T_{\rm c}$, 
where the distance between pions is still large \footnote{For instance, at $T=150\,{\rm MeV}$ 
the mean free path of pions is $\lambda \simeq 1/(n\;\sigma_{\pi \pi}^{\rm elastic}) \simeq 2\,{\rm fm}$.}, 
and where the correlation between pions is still small \cite{Correlation_5,Correlation_10}. 

According to {\it eq.}~(\ref{pion_mass_shift_T_1}) the pion mass increases with increasing temperature,  
in line with other theoretical investigations. Especially, more sophisticated
approaches yield very similar or even same results for the pion mass shift,
{\it e.g.} Chiral Perturbation Theory (ChPT) \cite{mass_shift2}, Nambu-Jona-Lasinio-model
\cite{mass_shift3}, QCD sum rules \cite{mass_shift4,mass_shift5}, Linear
Sigma Model \cite{mass_shift6,mass_shift7,mass_shift8,mass_shift9},
Mean Field Approximation \cite{mass_shift10} and  Virial Expansion
\cite{mass_shift11}. For instance, we can compare our findings 
(\ref{expression3}) - (\ref{pion_mass_shift_T_1}) with the results of ChPT,
since at order ${\cal O} (T^2)$ the pions are treated as free particles within ChPT, {\it e.g.} \cite{f_pi_T_5}.
For $\langle \hat{\overline q} \hat{q} \rangle_T$ \cite{Lit_5,qq_T_1}, for $f_{\pi} (T)$ \cite{f_pi_T_5} 
and for $m_{\pi}(T)$ \cite{GOR_T_3,f_pi_T_5}, the ChPT yields following expressions:
\bea
\langle \hat{\overline q} \hat{q} \rangle_T &=& \langle \hat{\overline q} \hat{q} \rangle_0
\left( 1 - \frac{1}{8} \frac{T^2}{f_{\pi}^2}\right) + {\cal O} \left(T^4 \right)\,,
\label{qq_T}
\\
\nonumber\\
f_{\pi} (T) &=& f_{\pi}
\left( 1 - \frac{1}{12} \frac{T^2}{f_{\pi}^2} \right) + {\cal O} \left(T^4 \right)\,,
\label{f_pi_T}
\\
\nonumber\\
m_{\pi} (T) &=& m_{\pi} \left( 1 + \frac{1}{48}\,\frac{T^2}{f_{\pi}^2} \right) + {\cal O} \left(T^4 \right)\;.
\label{pion_mass_shift_T_2}
\eea
The difference between our results given in {\it eqs.}~(\ref{expression3}) - (\ref{pion_mass_shift_T_1}) 
and the results of ChPT given in {\it eqs.}~(\ref{qq_T}) - (\ref{pion_mass_shift_T_2}) simply consists 
in the function $B_1$, reflecting the fact that according to the above mentioned references 
the ChPT considers a Bose gas and the limit $B_2 \rightarrow 1$. 
The result given in {\it eq.}~(\ref{qq_T}) has also been derived
in \cite{F_pi_4} and confirmed later on within the Sigma-model \cite{F_pi_1}.
The given result in {\it eq.}~(\ref{f_pi_T}) is confirmed,
{\it e.g.} in \cite{Ioffe1,F_pi_1,F_pi_2,F_pi_3}.  
In this respect we should refer the
interested reader to Ref.~\cite{F_pi_3} where a P\'ade approximation of
$f_{\pi} (T)$ has been
established, valid for arbitrary temperatures $T \le T_{\rm c}$. Nonetheless,
to apply that result $f_{\pi} (T)$ one needs a P\'ade
approximation for $\langle \hat{\overline q} \hat{q} \rangle_T$ as well
in order to be consistent within the entire framework. We also refer to
\cite{Dyson_Schwinger_1,Dyson_Schwinger_2} where a stronger decrease of the
pion decay constant $f_{\pi} (T)$ and a stronger increase of pion mass
$m_{\pi} (T)$ for high temperatures were found within the framework
of Dyson-Schwinger Equations. 

It is worth mentioning, that for in-medium pion mass shift there is an experimental 
result available \cite{Exp1,Exp2,Exp3,Exp4}, namely for the case of 
a deeply bound pionic state with a $\pi^{-}$ at finite baryonic density. 
Especially, a significant pion mass upshift of 
$19.3 \pm 2.7$ MeV in $^{207}{\rm Pb}$, {\it i.e.} approximatly at nuclear 
saturation density $n_0 \simeq 0.17\,{\rm fm}^{-3}$, has been found.

Another important experimental fact concerns the Constituent Quark Number 
scaling recently found \cite{CQS1,CQS2}. Such a scaling refers to an evident 
dependence of hadron elliptic flow $v_2$ on the number of constituent quarks 
in the hadron under consideration. This experimental fact can be understood by 
assuming that FO and pre-hadronization after QGP occurs at an intermediate 
constituent quark stage \cite{CQS3,CQS4}. Accordingly, the pre-hadrons 
are made of constituent quarks which have a mass of about $M_q \simeq 
(200 -300)\,{\rm MeV}$. For the pre-pions (this argumentation is not valid 
for all the other hadrons, because they 
are considerably heavier than the pions) in QGP it would imply a mass much 
higher than their vacuum mass $m_{\pi} \simeq 138\,{\rm MeV}$. From this point 
of view a mass increase with increasing temperature and baryon density is in 
agreement with these experimental facts. Indeed, QCD sum rules \cite{GOR_T_4} 
and QCD Dyson-Schwinger Equation \cite{Dyson_Schwinger_1,Dyson_Schwinger_2} 
predict a very strong pion mass up-shift near $T_{\rm c}$. In 
this respect we note, that a quark clustering like $qq, q\overline{q}, gg, qg$ 
etc. in the sQGP state has also been proposed in 
\cite{Clustering_1,Clustering_2}.

\section{Results and Discussion}\label{results_discussions}

In this Section we will present and discuss the solution of the coupled 
set of differential equations (\ref{DE_20}) and (\ref{DE_25}) for massive 
pions, once with the vacuum pion mass and once with a temperature dependent 
pion mass according to {\it eq.}~(\ref{GOR_10}). In order to ascertain the 
impact of pion mass itself on the FO process, we will compare these results 
with the case of massless pions. 

\begin{figure}[!ht]
\begin{center}
\hspace{-0.5cm}\includegraphics[angle=270,scale=0.5]{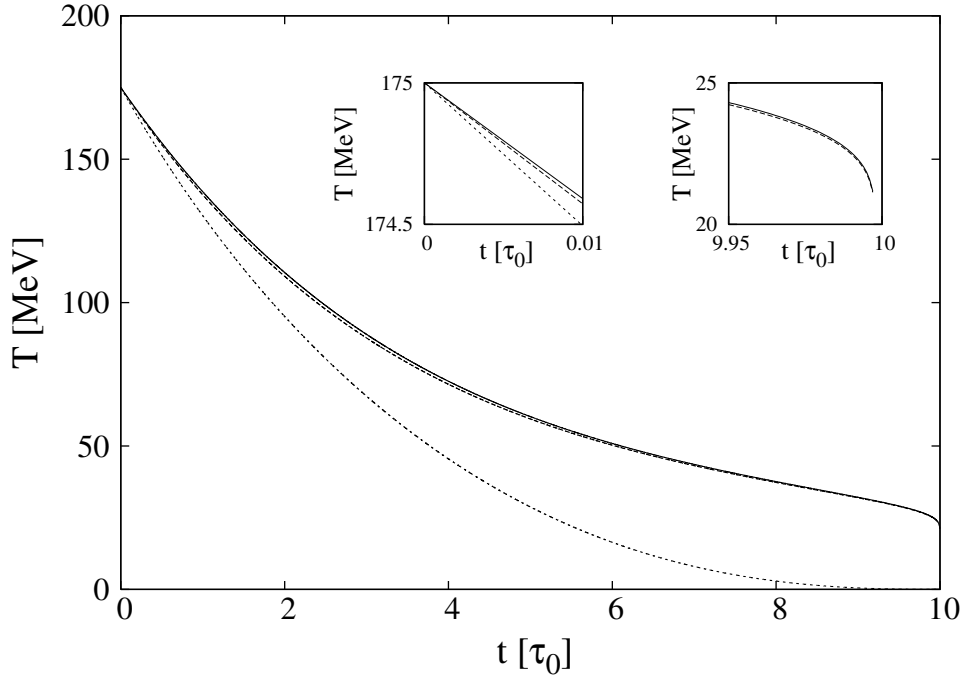}
\caption{Temperature of interacting component as function of time, when 
the matter crosses a finite Freeze Out layer. In the numerical evaluation we have taken 
$\tau_0 = 1\,{\rm fm}/c$ and $L = 10\,\tau_0$. Three situations of 
Freeze Out are considered: a pion gas with a temperature dependent pion mass 
$m_{\pi} (T)$ (solid line, top), the case of pions with a constant vacuum pion 
mass $m_{\pi}$ (dashed line, middle), and for massless pions $m_{\pi}=0$ 
(dotted line, bottom). As the system expands, the temperature of interacting 
component drecreases in all three cases, since the particles with larger 
momentum freeze out faster. The lighter the pions are, the 
faster the Freeze Out proceeds. In the left small figure the marginal impact 
of pion mass shift even at extremly high temperatures is shown. The right 
small figure shows the fast drecrease of temperature at the post Freeze Out 
surface. The substantial difference between the curves for massive pions and 
massless pions shows how important the impact of pion mass on the Freeze Out 
process is.}
\label{fig:Temperature}
\end{center}
\end{figure}

\begin{figure}[!ht]
\begin{center}
\hspace{-0.5cm}\includegraphics[angle=270,scale=0.5]{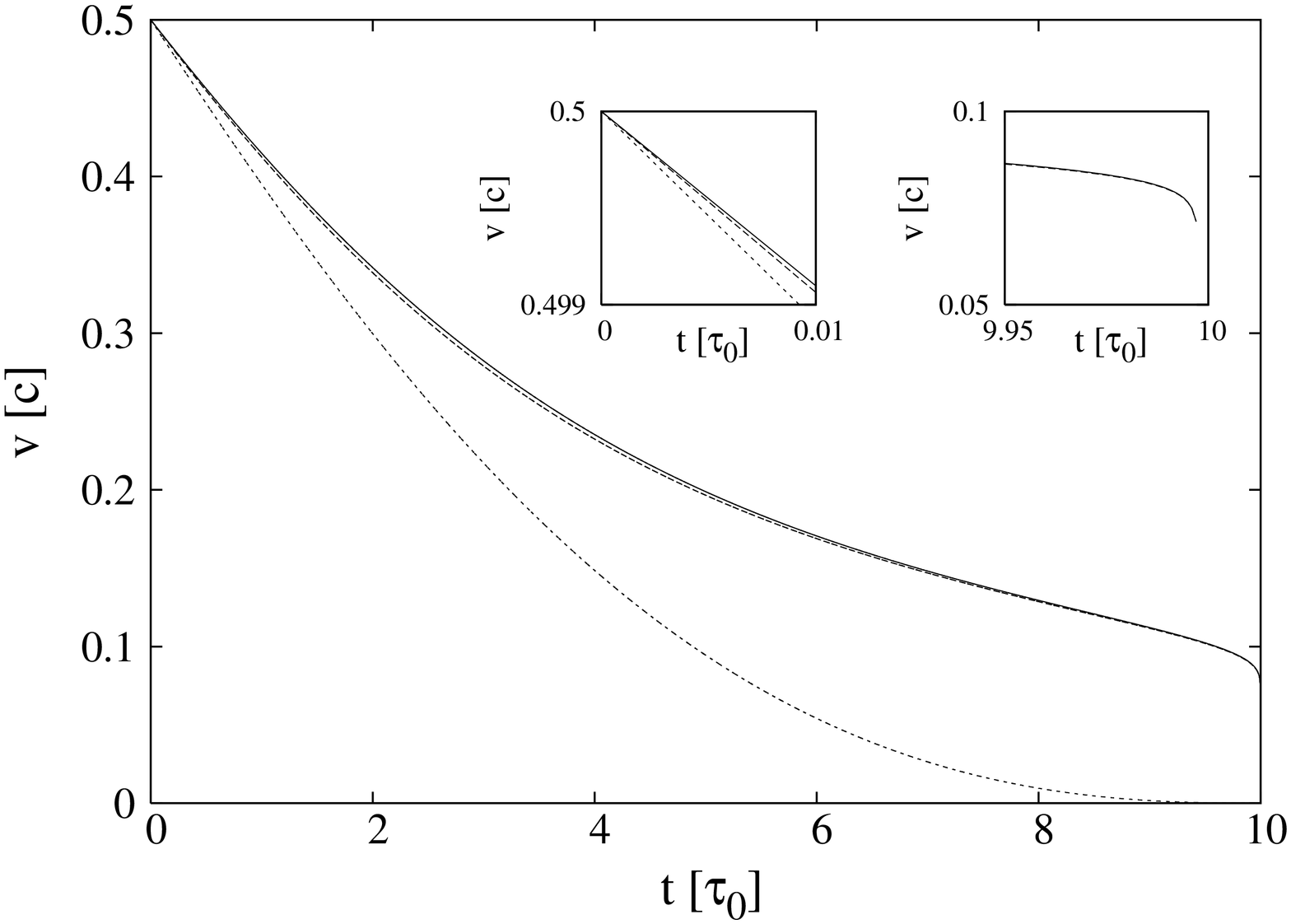}
\caption{Velocity of interacting component as function of time, when
the pions cross a finite Freeze Out layer. In the numerical evaluation 
we have taken $\tau_0 = 1\,{\rm fm}/c$ and
$L = 10\,\tau_0$. Three cases are considered:
massive pions with a temperature dependent pion mass $m_{\pi} (T)$
(solid line,top), massive pions with a constant vacuum pion mass $m_{\pi}$
(dashed line,middle), and massless pions $m_{\pi}=0$ (dotted line,bottom). 
The line of in-medium modified pion mass (solid line) is above the 
line of a constant pion mass. This is due to the fact that the 
pion mass is a bit larger in the former case. During the
Freeze Out process the fastest particles leave the interacting matter
component first, so the velocity of the remaining non-interacting
component decreases. The left small figure shows the tiny impact
of pion mass shift, while the right small figure shows the fast
dropping of velocity at the very end of the Freeze Out process.
The remarkable difference between the curves for massive pions and
massless pions shows the importance of pion mass on the Freeze Out process.}
\label{fig:Velocity}
\end{center}
\end{figure}

The differential equations (\ref{DE_20}) and (\ref{DE_25}) are solved with the 
aid of Runge-Kutta-Method \cite{Runge_Kutta}. The limit $m_{\pi} \rightarrow 0$ 
given in {\it eqs.}~(\ref{DE_26}) and (\ref{DE_27}) has been solved independently 
using MAPLE \cite{MAPLE}, {\it i.e.} used as independent verification. 
The two boundary conditions of these first-order differential equations are 
the initial temperature on pre-FO hypersurface, 
$T_{\rm pre-FO} \simeq T_{\rm c}=175\,{\rm MeV}$, and the initial flow 
velocity on pre-FO hypersurface of the finite layer, $v_{\rm pre-FO}=0.5\,c$. 
Both of these are typical values; note that such initial temperatures at small 
baryonic densities can be reached for instance at SPS \cite{Cleymans_Redlich}.
In the numerical evaluation we have taken $\tau_0 = 1\,{\rm fm}/c$ and 
$L = 10\,\tau_0$. 
The results are plotted in {\it figs.}~\ref{fig:Temperature} and \ref{fig:Velocity}.

\begin{figure}[!ht]
\begin{center}
\hspace{-0.5cm}\includegraphics[angle=270,scale=0.5]{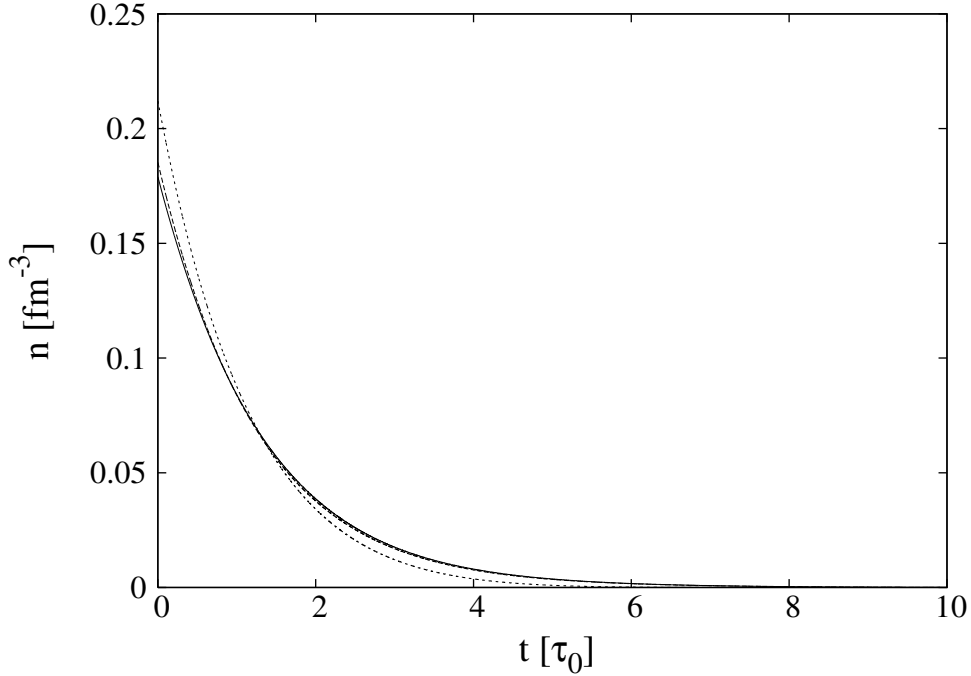}
\caption{Density of interacting component as function of time, when
the pions cross a finite Freeze Out layer. In the numerical evaluation 
we have taken $\tau_0 = 1\,{\rm fm}/c$ and $L = 10\,\tau_0$. Three cases 
are considered: massive pions with a temperature dependent pion mass $m_{\pi} (T)$
(solid line), massive pions with a constant vacuum pion mass $m_{\pi}$
(dashed line), and massless pions $m_{\pi}=0$ (dotted line). During the
Freeze Out process the particles leave the interacting matter
component, so the density of the remaining interacting
particles decreases. For the strong decrease of pion density we recall that 
for low temperatures the particles decreases exponentially: 
$\lim_{T \rightarrow 0} \; n (T) = g_{\pi} /(4 \pi^2) m_{\pi}^2 \, T \, 
\sqrt{2 \pi / a}\, {\rm exp} (-a)$; $a = m_{\pi}/T$.}
\label{fig:Density}
\end{center}
\end{figure}

Let us first consider the temperature of the interacting component plotted 
in {\it fig.}~\ref{fig:Temperature}. The temperature decreases in time, because 
the number of interacting particle decreases in time, 
see also {\it fig.}~\ref{fig:Density}. The difference between the case of a 
temperature dependent pion mass $m_{\pi}(T)$ (solid line) and the case of 
massive pions with constant vacuum mass $m_{\pi}$ (dashed line) is marginal. 
The reason for that can be understand by means of the small left figure of 
{\it fig.}~\ref{fig:Temperature}: At high temperatures $T \simeq T_{\rm c}$ the pion 
suffers a mass up-shift of about $5$ percent of it's vacuum mass, see {\it fig.}~\ref{fig:mass_shift}. 
And indeed, at such high temperatures 
we find a difference of about $5$ percent in the temperature slope 
between the case $m_{\pi} (T)$ and $m_{\pi}$. However, 
at such high temperatures the impact of a pion mass is small anyway, because 
the averaged momentum of pions due to thermal motion is considerably higher 
than the pion mass itself. 
At lower temperatures the impact of a pion mass on FO process is 
stronger because the averaged momentum of pions becomes comparable with the 
pion mass itself. However, already at moderate high temperatures of about 
$T \simeq 135\,{\rm MeV}$ the pion mass up-shift is less than $5$ 
percent, see {\it fig.}~\ref{fig:mass_shift}, and we cannot expect any longer 
a significant impact of the mass shift on the FO process. In fact, while the 
in-medium modification of pion 
mass shift has almost no impact, it turns out that there is a significant 
deviation between the case of massive pions and massless pions. This is due 
to the fact mentioned, that at moderate high temperatures the pion mass becomes 
very comparable to the averaged momentum of the pion. 
In the right small figure implemented in {\it fig.}~\ref{fig:Temperature} we 
can see the fast dropping of temperature at the end of the FO process. 
This is because of the factor $L/(L - t)$ in front of {\it eq.}~(\ref{DE_20}) 
which becomes infinite at the post-FO side of the hypersurface. 

Let us now consider the flow velocity of the interacting component plotted in 
{\it fig.}~\ref{fig:Velocity}. The flow velocity decreases in time, because the 
number of interacting particles also decreases, see {\it fig.}~\ref{fig:Density}. 
As in case of temperature, the difference between 
a FO scenario with a temperature dependent pion mass $m_{\pi}(T)$ (solid line) 
and a constant vacuum pion mass $m_{\pi}$ (dashed line) is negligible. The 
reason for such a behaviour is the very same as in case of temperature: At 
pre-FO side of the layer, there is a remarkable pion mass up-shift of 
about $5$ percent compared to vacuum pion mass, which implies roughly a $10$ 
percent impact on the flow velocity, as seen in the small left figure of 
{\it fig.}~\ref{fig:Velocity}. However, at such high temperatures 
$T \simeq T_{\rm c}$ the pion mass is considerably smaller than the averaged 
pion momentum due to thermal motion, so that the impact of a pion mass is 
negligible. As the system 
cools down, the pion mass approaches rapidly the vacuum pion mass, see 
{\it fig.}~\ref{fig:Temperature}. That means, at moderate temperatures of 
$T \simeq 135\,{\rm MeV}$ where a pion mass starts to have some impact of 
the FO process, the pion mass up-shift is already less than $5$ percent, so 
that the impact of in-medium modification of pion mass on FO process becomes 
negligible. However, there is a considerable difference between the case of 
massive pions and massless pions, as can be seen in {\it fig.}~\ref{fig:Velocity}.
The right small figure implemented in {\it fig.}~\ref{fig:Velocity} shows that 
at the post-FO side of hypersurface the velocity falls down rapidly 
because of the factor $L/(L-t)$ which becomes infinite at times near $L$.

The {\it fig.}~\ref{fig:Density} shows the rapid decrease of particle density 
of the interacting component. More than $90$ percent of the particles get 
frozen out before a time scale of $t \simeq 3 \,\tau_0$ is reached, 
both for massive and massless case. 

\begin{figure}[!ht]
\begin{center}
\hspace{-0.5cm}\includegraphics[angle=270,scale=0.25]{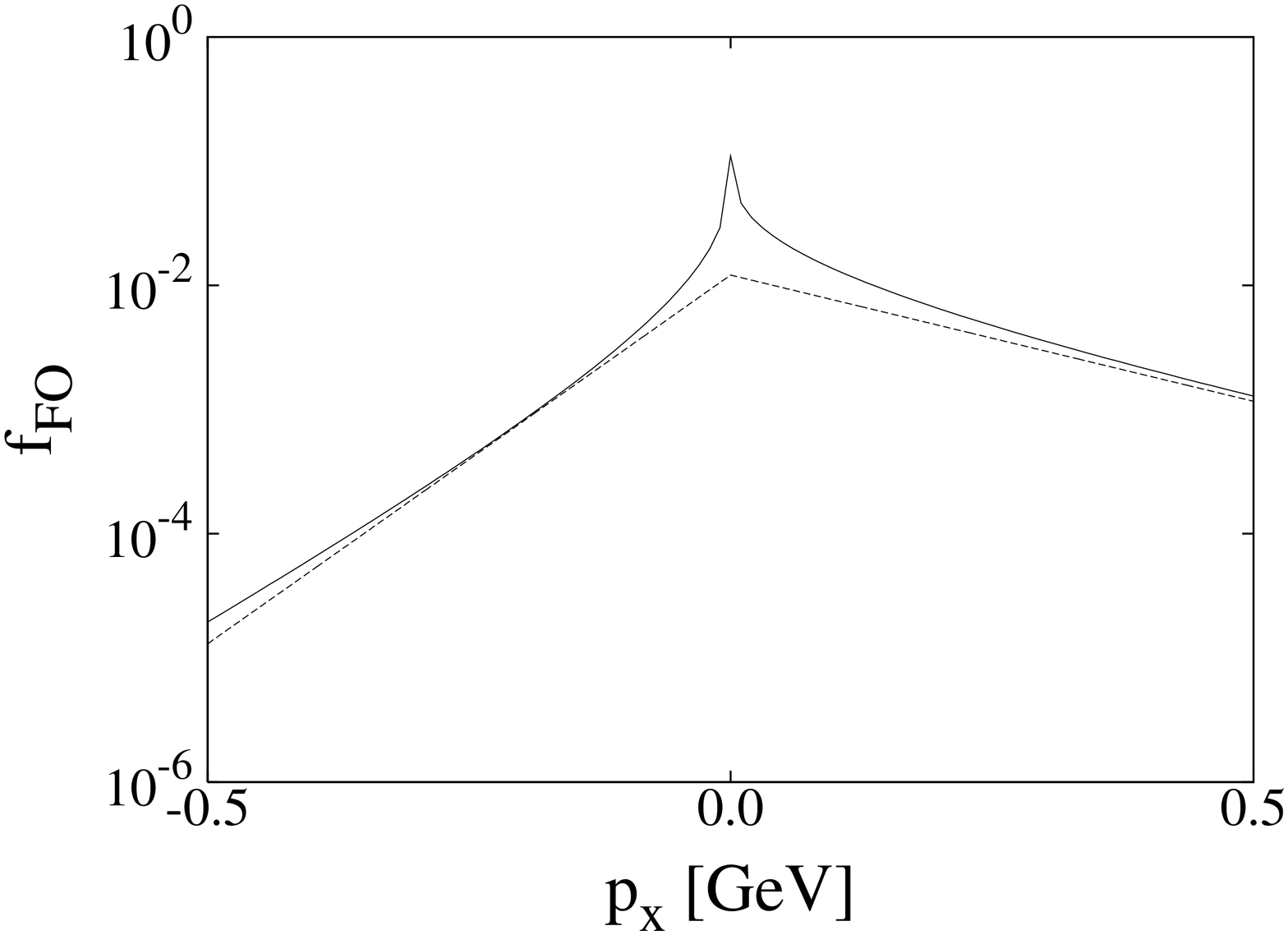}
\hspace{-0.5cm}\includegraphics[angle=270,scale=0.25]{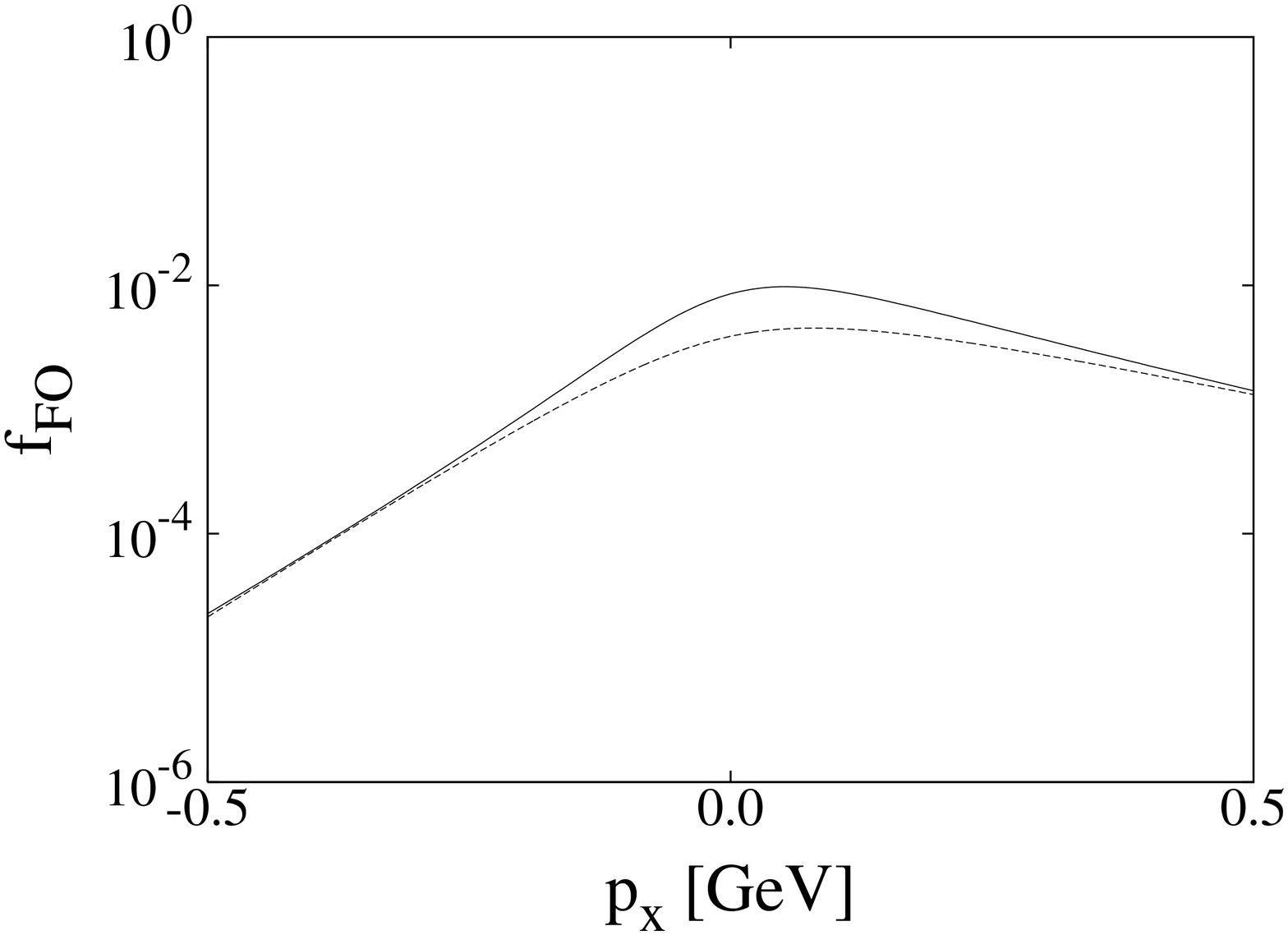}
\caption{{\bf Left pannel:} Massless pions. 
Solid line: FO distribution function $f_{\rm FO} (p_x) = f_f (t=L,p_x)$,
evaluated by means of {\it eq.}~(\ref{FO_70}) with $T(t)$ and $v(t)$ evaluated 
previously, see {\it figs.}~(\ref{fig:Temperature}) and (\ref{fig:Velocity}).
Dashed line: A fit of the above solid line by a thermal ({\it i.e.} J\"uttner) distribution with 
$T=125\,{\rm MeV}$ and $v=0.49\,c$. Both distribution functions peak sharply. 
At low momenta $p_x$, there is a considerable difference between a thermal 
distribution and the FO distribution function, {\it i.e.} the FO distribution 
function $f_{\rm FO}(p_x)$ evaluated within the kinetic model applied 
is obviously not a thermal distribution. This is actually one of the main 
results of the kinetic FO model. {\bf Right pannel:} Massive pions.
Solid line: FO distribution function $f_{\rm FO} (p_x) \equiv f_f (t=L,p_x)$ 
evaluated by means of {\it eq.}~(\ref{FO_70}) with $T(t)$ and $v(t)$ evaluated
previously, see {\it figs.}~(\ref{fig:Temperature}) and (\ref{fig:Velocity}), 
and with an in-medium pion mass $m_{\pi} (T)$.
Dashed line: A fit of the above solid line by a thermal J\"uttner distribution with 
$T=140\,{\rm MeV}$ and $v=0.4\,c$ and $m_{\pi}(T)$. There is a considerable 
difference between a thermal distribution and the FO distribution function 
for a massive pion gas at low momenta $p_x$, a result which is in agreement 
with the corresponding statement in the massless case. 
The distributions for massless and massive pions peak at different momenta 
$p_x$! This is expected to effect flow measurements. The sharp peak is smeared 
out after summing up for fluid elements with different pre-FO flow velocities.}
\label{fig:Distribution1}
\end{center}
\end{figure}

\begin{figure}[!ht]
\begin{center}
\hspace{-0.5cm}\includegraphics[angle=270,scale=0.35]{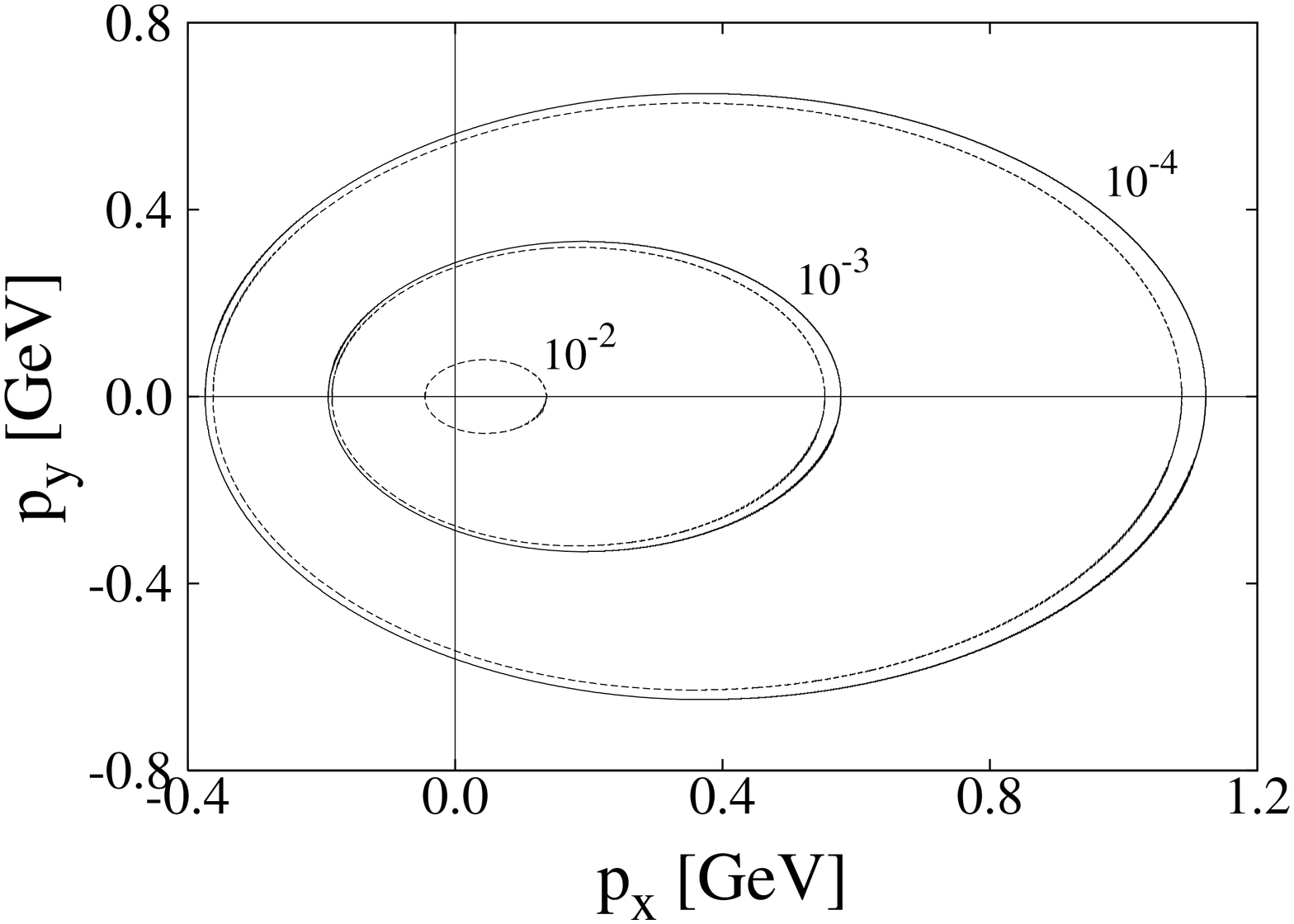}
\caption{Contur plot of FO distribution function, 
$f_{\rm FO} (p_x, p_y) \equiv f_f (t=L,p_x, p_y)$, evaluated by means of 
{\it eq.}~(\ref{FO_70}) with $T(t)$ and $v(t)$ evaluated previously, see 
{\it figs.}~(\ref{fig:Temperature}) and (\ref{fig:Velocity}). Solid line for the 
case of massive pions $m_{\pi}(T)$, Dashed Line for the case of massless pions. 
There is no curve for the case of massive pions if $f_{\rm FO} = 10^{-2}$ 
because the FO distribution function in case of massive pions is smaller 
than this value, see {\it fig.}~\ref{fig:Distribution1}. It illustrates the 
importance of pion mass for an accurate FO description. 
The non-thermal asymmetry of the post-FO particles is a strong and 
dominant feature, even for a time-like normal FO hypersurface or layer.}
\label{fig:Distribution2}
\end{center}
\end{figure}

In heavy-ion collision experiments the relevant experimental parameters 
are of course not the temperature $T (t)$ or flow-velocity $v(t)$ 
of interacting component.
Instead, via the experimentally accessible spectrum of the particles 
$dN/d^3 p$ one measures the post-FO distribution, which is formed in the 
applied model at the outer hypersurface, $f_{\rm FO} (p_x) \equiv 
f_f (t=L,p_x)$. The Cooper-Frye formula \cite{Cooper_Frye}
\bea
p_0\,\frac{d N}{d^3 p} &=& \int_{\sigma} d \hat{\sigma}_{\mu}
\,p^{\mu}\,f_{\rm FO} (p) 
\nonumber\\
&=& \int d V\,p_0\,f_{\rm FO} (p) \sim p_0\,f_{\rm FO} (p)
\label{Cooper_Frye}
\eea
allows us to calculate the particle spectrum 
($d \hat{\sigma}_{\mu} = (d V, 0,0,0)$ is the time-like normal vector of an 
infinitesimal element of the post-FO hypersurface). In the one-dimensional 
model used, the volume $V$ of the pionic fireball is ill-defined 
\footnote{Since the volume of the pionic fireball is not defined, the global energy and momentum 
conservation cannot be applied. However, global conservation laws can lead to different 
results if and only if the curvature of the surface is large, thus leading to a significant 
divergence. Such a situation has been considered in \cite{Magas}, where also  
related issues concerning an expanding system are discussed.}, 
so that only a 
qualitative comparison with experimental data, for instance the transverse 
spectrum, is possible, {\it i.e.} $\frac{d N}{d p_x} \sim f_{\rm FO} (p_x)$, where
$p_x$ is locally pointing out in transverse direction. Due to the 
marginal impact of in-medium pion mass modification it is meaninful 
to compare the case of massive pions $m_{\pi}(T)$ with the case of 
massless pions, and both of each with a simple thermal distribution. 

In {\it fig.}~\ref{fig:Distribution1} the frozen out distribution functions 
$f_{\rm FO} (p_x)$ for massless and massive pions are plotted and compared 
with a thermal distribution. It is shown, that at low momenta the distribution 
function $f_{\rm FO} (p_x)$ evaluated within the kinetic FO model differs 
considerably from a simple thermal distribution, both for massless and 
massive pions. This confirms a very recent FO evaluation where a gradual 
FO with Bjorken expansion and with a constant pion mass 
$m_{\pi} = 0.138\,{\rm MeV}$ has been considered \cite{V_Magas1,V_Magas2}. 
Notice, that the peak position of distribution function for the case of 
massive pions differs from the peak position for a massless pion gas. This 
is expected to effect experimental flow measurements. 

In {\it fig.}~\ref{fig:Distribution2} the contur plot $f_{\rm FO} (p_x,p_y)$ for the 
massive case $m_{\pi}(T)$ and the massless case $m_{\pi}=0$ is compared. 
For large momenta $p$ there is only a slight difference between 
the FO of massive and massless pions. However, at low $p$ there is a 
significant deviation between a massive pion gas and the massless case. 
The {\it figs.}~\ref{fig:Distribution1} and \ref{fig:Distribution2} demonstrate 
that the mass of pions is relevant for an accurate FO description.

\section{Summary}\label{summary}

The strong interacting Quark Gluon Plasma is not directly accessible via 
experiments. In fact, one has to trace back from the experimental data to 
the initially formed new state of matter during an ultra-relativistic 
heavy-ion collision. This implies a detailed and accurate description of all 
subsequent stages of heavy-ion collision process. Within our FO model of a pion gas 
described by a Boltzmann gas, we have considered the problem how strong the impact 
of an in-medium modification of the pion mass on the kinetic Freeze Out process is. 

We have considered the kinetic FO process through a finite layer for a pion gas with 
finite pion mass and with massless pions. Throughout the investigation, all calculations 
have been consistently performed within the model of a pion gas with elastic scatterings 
among them and approximated by a Boltzmann gas.

First, we have found, that there 
is a strong impact of the finite pion mass on the FO process compared to a 
FO process with massless pions. This result is highlighted in 
{\it figs.}~\ref{fig:Temperature} and \ref{fig:Velocity}, where a significant 
modification of the basic thermodynamical functions temperature $T(t)$ 
and flow velocity $v(t)$ of elastic interacting component of pion gas inside 
a finite FO layer has been found. 

Hereafter, we have investigated how strong the impact of in-medium 
modification of pion mass on the FO process is. To determine the
pion mass at finite temperature, the generalized GOR relation at finite temperature
has been applied as a given fact, and the needed expressions $\langle \hat{\overline q} \hat{q} \rangle_T,
f{\pi} (T)$ have been evaluated by a thermal average over one-pion states. 
The needed pion matrix elements were evaluated by the soft-pion theorem. Especially, the
modification of effective pion mass is increasing with increasing
temperature, becoming significant around $100\,{\rm MeV}$ and reaches
$\simeq 5\,{\rm MeV}$ at $T \simeq 180\,{\rm MeV}$, see {\it fig.}~\ref{fig:mass_shift}.

It has turned out, that the impact of in-medium pion mass shift 
$m_{\pi}(T)$ on the FO process is marginal compared to a massive pion gas 
with a constant pion mass $m_{\pi}$. The physical reason for that can be 
understood as follows: A significant pion mass shift occurs at high 
temperatures $T \simeq T_{\rm c}$. However, at such high temperatures the 
impact of the pion mass itself on FO process is negligible because 
the momentum of thermal motion is sizebly larger than the pion mass. 
During the FO process the temperature decreases rapidly, 
and after the initial $\sim 2\,\tau_0$ time it was below 
$T\simeq 100\,{\rm MeV}$. But already at moderate temperatures 
$T \le 135\,{\rm MeV}$, where the impact of pion mass on FO process becomes 
relevant, there is no longer a remarkable in-medium modification of pion mass. 

For a qualitative comparison with experimantal data and in order to 
investigate the importance of a finite pion mass for the FO process, 
in {\it figs.}~\ref{fig:Distribution1} the FO distribution function 
$f_{\rm FO} (p_x)$ has been plotted, both for massive $m_{\pi}(T)$ and 
massless $m_{\pi}=0$ pions. The FO distribution functions have been 
compared with a thermal distribution. It is shown that at low momenta there 
is a considerable difference to a thermal spectrum, both for a massive 
and massless pion gas. The distributions peak at different momenta $p_x$ 
which is expected to effect experimental flow measurements. 
The contur plot in {\it fig.}~\ref{fig:Distribution2} elucidates 
a remarkable difference between massive and massless pions at low momenta.
Both {\it figs.}~\ref{fig:Distribution1} and \ref{fig:Distribution2} show the 
importance of a finite pion mass for an accurate description of the FO process.

The numerical order of in-medium pion mass shift at finite temperatures used 
in our model calculation is in some sense also encouraged from experiments 
which have found a pion mass up-shift of $19.3 \pm 2.7\,{\rm MeV}$ around 
baryon saturation density $n_0 \simeq 0.17 \,{\rm fm}^{-3}$ 
\cite{Exp1,Exp2,Exp3,Exp4}.
However, we have to take into account, that the in-medium pion mass 
shift might be much more pronounced at high temperatures 
$T \simeq T_{\rm c}$ than used in our model calculations. For instance, 
QCD sum rules \cite{GOR_T_4} and QCD Dyson-Schwinger Equation  
\cite{Dyson_Schwinger_1,Dyson_Schwinger_2} predict a strong pion mass 
up-shift near $T_{\rm c}$. 

Such a strong in-medium modification 
of pion mass would be also in line with the argumentation of quark 
pre-clustering inside the QGP for temperatures higher but near the QCD 
phase transition at $T \ge T_{\rm c}$, because the quarks will have a 
constituent quark mass of about $M_q \simeq 200 - 300\,{\rm MeV}$, 
so that the mass 
of pre-pions for such temperature regions would be much higher than the vacuum 
pion mass. Our results indicate that an earlier pre-hadronization and 
FO at higher temperatures would increase the sensitivity on the changing 
of effective pion mass.

From experimental side, as mentioned in Section \ref{mass_shift}, such a 
conclusion is mainly supported by the experimental fact of Constituent Quark 
Number scaling, {\it i.e.} the dependence of hadron elliptic flow $v_2$ 
from the number of constituent quarks of the hadrons. Therefore, for a more 
comprehensive analysis how strong the impact of in-medium modification of the 
pion mass for an accurate description of a heavy-ion collision process is, 
further insides of (pre-)pion mass shift near 
(and beyond) the critical temperature $T_{\rm c}$, both from theoretical as 
well as experimental side, are mandatory. Especially, the pion mass modification caused 
by non-elastic scatterings among the pions have to be evaluated in order to 
describe the chemical FO process. 

In summary, in our model study we come to the conclusion that, while a pion mass $m_{\pi}$ 
has a significant impact on the kinetic FO process, the temperature dependence of the pion mass
$m_{\pi}(T)$ is negligible for an accurate description of the kinetic FO process. 

\section*{Acknowledgements}  
S.Z. thanks for the pleasant hospitality at the Bergen Center for 
Computational Science (BCCS) and Bergen Center Physics Laboratory (BCPL) 
at the University of Bergen/Norway. Kalliopi Kanaki is acknowledged for 
giving valuable information and suggestions. The authors acknowledge 
detailed and fruitful discussions with Volodymyr Magas. Dag Toppe Larsen, 
Magne H\r{a}v\r{a}g and Miklos Zetenyi are acknowledged for their computer assistance. 
Sincere thanks to Fred Kurtz for a critical reading of the manuscript.

\section*{Appendix A}

The function $G_n^{\pm}\; (n=1,2)$ is defined by 
\bea
G_n^{\pm} (m_{\pi}, v, T) &=& \frac{1}{T^{n+2}} \, \int\limits_0^{\infty}
d p \; p \left( \sqrt{p^2 + m_{\pi}^2} \right)^n \; {\rm E}_1
\left(\frac{\gamma}{T} \, \sqrt{p^2 + m_{\pi}^2} \pm \frac{\gamma\, v\, p}
{T}\right)\;,
\label{appendix_5}
\eea
where ${\rm E}_1$ is a special case of incomplete Gamma-function
\cite{Abramowitz_Stegun} 
\bea
{\rm E}_1 (x) &=& \int \limits_{x}^{\infty} dt \; t^{-1} \;
{\rm e}^{- t}\;.
\label{appendix_10}
\eea
In the massless case we get the temperature independent limits 
\bea
\lim_{m_{\pi \rightarrow 0}}\,
G_1^{\pm} &=& \frac{2}{3}\, \frac{1}{\gamma^3} \, \frac{1}{(1 \pm v)^3}\,,
\nonumber\\
\lim_{m_{\pi \rightarrow 0}}\,
G_2^{\pm} &=& \frac{3}{2}\, \frac{1}{\gamma^4} \, \frac{1}{(1 \pm v)^4}\,.
\eea
The function $K_n$ is the Bessel function of second kind
\cite{Abramowitz_Stegun}, defined by 
\bea
K_n (z) = \frac{2^n\; n!}{(2 n)!} z^{-n} \int_z^{\infty} d x \;
{\rm e}^{-x} \; (x^2 - z^2)^{n - 1/2}\;.
\label{appendix_15}
\eea

\section*{Appendix B}

In this Appendix we will evaluate the matrix elements in {\it eqs.}~(\ref{Gibbs_15}) and (\ref{Gibbs_20}) 
with the aid of soft-pion theorem \cite{F_pi_5,soft1,soft2,soft3,Toki} (a derivation of soft-pion theorem
for the here used simplier case of "free" pions can be found in Appendix of Ref.~\cite{zschocke}) given by
\begin{eqnarray}
\lim_{\atop p_2 \to 0} \lim_{\atop p_1 \to 0}
\langle \pi^m (p_2) | \hat{\cal O} (x) | \pi^n (p_1) \rangle
&=& \frac{1}{f_{\pi}^2} \langle 0 | \Bigg[ \hat{Q}_A^m ,
\bigg[ \hat{\cal O} (x) \; , \; \hat{Q}_A^{n\;\dagger}
\bigg]_{-} \Bigg]_{-} |0\rangle\;,
\label{appendixB_5}
\end{eqnarray}
where $\hat{Q}_A^n$ is the (time independent) axial charge, 
{\it i.e.} the spatial integral over zeroth component of axial vector current (\ref{axial_current}):
$\hat{Q}_A^n = \int d^3 {\bf r} \; \hat{A}_0^n ({\bf r}, t)$, and $\hat{Q}_A^{n\;\dagger} = \hat{Q}_A^n$; 
and $n,m = 1,2,3$ are the isospin indices.
$[\hat{A},\hat{B}]_{-}=\hat{A}\hat{B} - \hat{B}\hat{A}$ is the
commutator. Here, we are interested in pion matrix elements 
of the following two-quark operators at $x=0$:
\bea
\hat{\cal O}_1 &=& \hat{\overline q} \hat{q}
\equiv \frac{1}{2} \sum\limits_{{\rm i} = 1}^3
\sum\limits_{\alpha,\beta = 1}^4
(\gamma_0)_{\alpha \beta}\left(
\hat{u}^{{\rm }i\;\dagger}_{\alpha} \, \hat{u}_{\beta}^{\rm i} + 
\hat{d}^{{\rm i}\;\dagger}_{\alpha} \, \hat{d}_{\beta}^{\rm i}\right)\;,
\label{appendixB_10}
\\
\nonumber\\
\hat{\cal O}_2 &=& \hat{A}_{\mu}^3 = 
\frac{1}{2}\; 
\left(\hat{\overline u} \gamma_{\mu} \gamma_5 \hat{u} 
 - \hat{\overline d} \gamma_{\mu}\,\gamma_5 \hat{d}\right)
\equiv \frac{1}{2} \sum\limits_{{\rm i} = 1}^3
\sum\limits_{\alpha,\beta = 1}^4\;
(\gamma_0 \gamma_{\mu}\gamma_5)_{\alpha \beta}\;
\left(\hat{u}^{{\rm i}\;\dagger}_{\alpha} \, \hat{u}_{\beta}^{\rm i} - 
\hat{d}^{{\rm i}\;\dagger}_{\alpha} \, \hat{d}_{\beta}^{\rm i} \right)\;,
\label{appendixB_11}
\eea
where the greek letters $\alpha, \beta$ denote Dirac indices, and
${\rm i}$ is the color index of quarkfields. 
The equal-time anti-commutators of the full QCD quark fields 
read for quark operators of the same flavor
\bea
[\hat{q}^{\rm i}_{\alpha} ({\bf r}_1, t) \, , \,
\hat{q}^{{\rm j}\;\dagger}_{\beta} ({\bf r}_2, t)]_{+}
= \delta^{(3)} ({\bf r}_1 - {\bf r}_2) \; \delta_{\alpha \beta}
\;\delta^{{\rm i}\,{\rm j}}\;,
\label{appendixB_15}
\eea
where $[\hat{A},\hat{B}]_{+}=\hat{A}\hat{B} + \hat{B}\hat{A}$ is the
anti-commutator, while quark fields of different flavor anti-commute. 
Using identities like 
$[\hat{A},\hat{B} \hat{C}]_{-} = [\hat{A},\hat{B}]_{+} \,\hat{C} - \hat{B} [\hat{A},\hat{C}]_{+}$
and the anti-commutator relation (\ref{appendixB_15}) we obtain 
\bea
\lim_{\atop p_2 \to 0} \lim_{\atop p_1 \to 0} 
\langle \pi^{m} (p_2) | \hat{\overline q} \hat{q} | \pi^{n} (p_1) \rangle &=& 
- \frac{1}{f_{\pi}^2} \langle 0 | \hat{\overline q} \hat{q} | 0 \rangle \;\delta^{m n} \,,
\label{matrixelement_1}
\\
\nonumber\\
\lim_{\atop p_2 \to 0} \lim_{\atop p_1 \to 0} 
\langle \pi^{1} (p_2) | \hat{A}_{\mu}^3 | \pi^3 (q) \pi^{1} (p_1) \rangle &=& 
- \frac{1}{f_{\pi}^2} 
\langle 0 | \hat{A}_{\mu}^3 | \pi^3 (q) \rangle = i\,\frac{1}{f_{\pi}} \, q_{\mu} \,,
\label{matrixelement_2}
\\
\lim_{\atop p_2 \to 0} \lim_{\atop p_1 \to 0}
\langle \pi^{2} (p_2) | \hat{A}_{\mu}^3 | \pi^3 (q) \pi^{2} (p_1) \rangle &=&
- \frac{1}{f_{\pi}^2}
\langle 0 | \hat{A}_{\mu}^3 | \pi^3 (q) \rangle = i\,\frac{1}{f_{\pi}} \, q_{\mu} \,,
\label{matrixelement_3}
\\
\nonumber\\
\lim_{\atop p_2 \to 0} \lim_{\atop p_1 \to 0}
\langle \pi^3 (p_2) | \hat{A}_{\mu}^3 | \pi^3 (q) \pi^3 (p_1) \rangle &=& 0\;,
\label{matrixelement_4}
\eea
where in the last step of {\it eqs.}~(\ref{matrixelement_2}) and (\ref{matrixelement_3}) 
we have used relation (\ref{pion_decay_constant_5}). 

\section*{Appendix C}

In our model we have taken into account the Gibbs average over one-pion 
states of an operator $\hat{\cal O}$, see {\it eq.}~(\ref{Gibbs_10}). Here we will consider the 
next order, {\it i.e.} the Gibbs average over two-pion states of an operator $\hat{\cal O}$:  
\bea
{\cal O} \left(T^4 \right) &=& \frac{1}{2!} \, 
\sum \limits_{n,m=1}^{3}\;\int \frac{d^3 p_1}{(2 \pi)^3\;2\,p_1^0}\;
\int \frac{d^3 p_2}{(2 \pi)^3\;2\,p_2^0}\;{\rm exp} \left(- \frac{p_1^0}{T} \right)\;
{\rm exp} \left(- \frac{p_2^0}{T} \right)
\nonumber\\
\nonumber\\
&& \times \langle \pi^m (p_2)  \pi^n (p_1)| \hat{O} | \pi^n (p_1)  \pi^m (p_2) \rangle\,,
\label{appendixC_5}
\eea
where the factor $(2!)^{-1}$ in front of (\ref{appendixC_5}) circumvents a double counting by 
integrating over the permutations $p_1 \leftrightarrow p_2$. 
With the aid of soft-pion theorem we obtain the pion matrix elements for chiral condensate 
and pion decay constant:
\bea
\sum \limits_{n,m=1}^{3} \lim_{\atop p_2 \to 0} \lim_{\atop p_1 \to 0} 
\langle \pi^m (p_2)  \pi^n (p_1)| \hat{\overline q} \hat{q} | \pi^n (p_1)  \pi^m (p_2) \rangle 
&=& \frac{9}{f_{\pi}^4}\;\langle 0 | \hat{\overline q} \hat{q} | 0 \rangle\;, 
\label{appendixC_10}
\\
\nonumber\\
\sum \limits_{n,m=1}^{3} \lim_{\atop p_2 \to 0} \lim_{\atop p_1 \to 0}
\langle \pi^m (p_2)  \pi^n (p_1)| \hat{A}_{\mu}^3 | \pi^3 (q) \pi^n (p_1)  \pi^m (p_2) \rangle
&=& \frac{4}{f_{\pi}^4}\,\langle 0 | \hat{A}_{\mu}^3 | \pi^3 (q) \rangle = - i \frac{4}{f_{\pi}^3}\,q_{\mu}\,,
\nonumber\\
\label{appendixC_15}
\eea
where in the last line we have applied {\it eq.}~(\ref{pion_decay_constant_5}). 
By means of these expressions we obtain 
\bea
\langle \hat{\overline q} \hat{q} \rangle_T &=& \langle \hat{\overline q} \hat{q} \rangle_0
\left( 1 - \frac{1}{8} \frac{T^2}{f_{\pi}^2}\; B_1 
+ \frac{1}{128} \frac{T^4}{f_{\pi}^4}\; B_1^2 \right) + {\cal O} (T^6) \,,
\label{appendixC_20}
\\
\nonumber\\
f_{\pi} (T) &=& f_{\pi}
\left( 1 - \frac{1}{12} \frac{T^2}{f_{\pi}^2} \; B_1 
+ \frac{1}{288} \frac{T^4}{f_{\pi}^4} \; B_1^2 \right) + {\cal O} (T^6) \,,
\label{appendixC_25}
\\
\nonumber\\
m_{\pi} (T) &=& m_{\pi} \left( 1 + \frac{1}{48} \frac{T^2}{f_{\pi}^2} \; B_1 
+ \frac{1}{4608} \frac{T^4}{f_{\pi}^4} \; B_1^2 \right) + {\cal O} (T^6) \,,
\label{appendixC_26}
\eea
and $B_1 \equiv B_1 (m_{\pi}/T)$.
A numerical evaluation shows that the terms of order ${\cal O} (T^4)$ 
contribute at most a few percent compared to the terms of order ${\cal O} (T^2)$ 
in the temperature region we are interested. 
Within ChPT approach very similar statements are found, but it should be mentioned that in ChPT 
non-elastic particle interactions yield additional contributions to order ${\cal O} (T^4)$, while 
{\it eqs.}~(\ref{appendixC_20}) - (\ref{appendixC_26}) are the results for a pion gas with elastic particle collisions 
only. The dynamical non-elastic interactions among the pions change not only the given numerical values in 
{\it eqs.}~(\ref{appendixC_20}) - (\ref{appendixC_26}) but also the sign of the coefficients in front of the 
given order ${\cal O} (T^4)$, {\it {\it e.g.}} {\it Ref.}~\cite{Ioffe2}, where the result (\ref{appendixC_20}) 
is also mentioned. But we note again, that in ChPT the given coefficients of the order ${\cal O} (T^2)$ 
remain untouched even when taking into account the contributions of non-elastic pion scatterings.

Furthermore, in a pion gas with elastic interactions and to all orders in temperature we obtain by iteration  
\bea
\langle \hat{\overline q} \hat{q} \rangle_T &=& \langle \hat{\overline q} \hat{q} \rangle_0
\left( 1 - \frac{1}{8}\,\frac{T^2}{f_{\pi}^2} +  
\sum \limits_{n=2}^{\infty} (-1)^n\,\frac{1}{n!} \,\frac{1}{8^{n}}\,\frac{T^{2 n}}{f_{\pi}^{2 n}}\,
B_1^{n} \right) \,,
\label{appendixC_30}
\\
\nonumber\\
f_{\pi} (T) &=& f_{\pi} \left( 1 - \frac{1}{12}\,\frac{T^2}{f_{\pi}^2} + 
\sum \limits_{n=2}^{\infty} (-1)^n\,\frac{1}{n!} \,\frac{1}{12^{n}}\,
\frac{T^{2 n}}{f_{\pi}^{2 n}}\,B_1^{n} \right) \,,
\label{appendixC_35}
\\
\nonumber\\
m_{\pi} (T) &=& m_{\pi}\left( 1 + \frac{1}{48}\,\frac{T^2}{f_{\pi}^2} +
 \sum \limits_{n=2}^{\infty} \,\frac{1}{n!} \,\frac{1}{48^{n}}\, 
\frac{T^{2 n}}{f_{\pi}^{2 n}}\,B_1^{n} \right)  \,,
\label{appendixC_40}
\eea
that means all higher orders are factorial suppressed even at temperatures near $ T_{\rm c}$.
Again we underline, that {\it eqs.} (\ref{appendixC_30}) - (\ref{appendixC_40}) are useful to study the impact of pion mass shift on kinetic FO of a pion gas (cease of elastic interactions 
among the pions). However, they are not applicable for modelling the pion mass shift impact on chemical FO process 
of a pion gas (cease of non-elastic interactions among the pions), because then the non-elastic interactions 
among the pions will change significantly the given series of higher order.

\end{document}